\documentclass[12pt, letterpaper]{article}

\usepackage{amsmath,amssymb}
\usepackage{cite}
\usepackage{fancyhdr}
\usepackage[top=1in, bottom=1in, left=1in, right=1in]{geometry}
\usepackage[dvipdfm, dvips]{graphicx}
\usepackage{hyperref}

\numberwithin{equation}{section}

\begin{document}


\setcounter{page}{0}
\date{}

\lhead{}\chead{}\rhead{\footnotesize{RUNHETC-2010-16\\SCIPP-10-11}}\lfoot{}\cfoot{}\rfoot{}

\title{\textbf{TASI Lectures on Holographic Space-time, SUSY, and Gravitational
Effective Field Theory\vspace{0.4cm}}}

\author{Tom Banks$^{1,2}$\vspace{0.7cm}\\
{\normalsize{$^1$NHETC and Department of Physics and Astronomy, Rutgers University,}}\\
{\normalsize{Piscataway, NJ 08854-8019, USA}}\vspace{0.2cm}\\
{\normalsize{$^2$SCIPP and Department of Physics, University of California,}}\\
{\normalsize{Santa Cruz, CA 95064-1077, USA}}}

\maketitle
\thispagestyle{fancy}

\begin{abstract}
\normalsize \noindent I argue that the conventional field theoretic
notion of vacuum state is not valid in quantum gravity. The
arguments use gravitational effective field theory, as well as
results from string theory, particularly the AdS/CFT correspondence.
Different solutions of the same low energy gravitational field
equations correspond to different quantum systems, rather than
different states in the same system. I then introduce {\it
holographic space-time} a quasi-local quantum mechanical
construction based on the holographic principle. I argue that models
of quantum gravity in asymptotically flat space-time will be exactly
super-Poincare invariant, because the natural variables of
holographic space-time for such a system, are the degrees of freedom
of massless superparticles.  The formalism leads to a non-singular
quantum Big Bang cosmology, in which the asymptotic future is
required to be a de Sitter space, with cosmological constant (c.c.)
determined by cosmological initial conditions. It is also
approximately SUSic in the future, with the gravitino mass $K
\Lambda^{1/4}$.

\end{abstract}


\newpage
\tableofcontents
\vspace{1cm}


\section{Vacuum states in non-gravitational quantum field theory}

QFTs in fixed space-time backgrounds, like Minkowski space, often
exhibit the phenomena of degenerate and/or meta-stable vacuum
states. In the semi-classical approximation these are solutions of
the field equations that preserve all the isometries of the
background, and for which there are no exponentially growing small
fluctuations. Typically, this requires the model to contain
fundamental scalar fields.  The potential energy density is a
function of these scalars, and multiple solutions occur when this
function has multiple minima.

In the semi-classical approximation, this is evidence for multiple
{\it superselection sectors} of the QFT:  the Hilbert space breaks
up into a direct sum of spaces, each associated with a different
minimum.  In the infinite volume limit, transitions between sectors
vanish because the Hamiltonian is an integral of a local energy
density.   Actually, this is only true in perturbation theory around
the true minima.  When non-perturbative physics is taken into
account, there are generally bubble nucleation processes, which
signal an instability of all but the lowest energy minima.
Superselection sectors only exist for minima which are exactly
degenerate, including all quantum corrections to the energy (the
energy differences between semi-classical vacua do not suffer from
renormalization ambiguities).

A more non-perturbative view of these phenomena is afforded by the
Wilsonian definition of quantum field theory.  A general QFT is
defined by a relevant perturbation of a CFT.  CFT's in turn are
defined by their spectrum of conformal primary operators and their
operator product expansions (OPEs).  In particular, this includes a
list of all the relevant operators, which might be added as
perturbations of the CFT, using the GellMann-Low formula to compute
the perturbed Green's functions.  The OPE allows us to perform these
computations.  Although there is no general proof, it is believed
that these conformal perturbation expansions are convergent in
finite volume.

The CFT has a unique conformally invariant vacuum state, which is
the lowest energy state if the theory is unitary.  However, in the
infinite volume limit the Hilbert space of the perturbed theory
might again separate into superselection sectors.  It might
also/instead have meta-stable states, but meta-stability always
depends on the existence of a small dimensionless parameter, the
life-time of the meta-stable state in units of the typical time
scale in the model.  In most explicit examples, this parameter is a
semi-classical expansion parameter for at least some of the fields
in the theory.

The following general properties of degenerate and meta-stable vacua
in QFT, follow from these principles:

\begin{itemize}

\item The short distance behavior of Green's functions, and the
high temperature behavior of the partition function of the theory
are independent of the superselection sector. Both are controlled by
the CFT.  The partition function in finite volume $V$ has the
asymptotic form

$$Z = e^{- c V^{\frac{2d - 1}{d}} E^{\frac{d - 1}{d}}} ,$$ where $d$
is the space-time dimension and $E$ the total energy.  This follows
from scale invariance and extensivity of the energy. Extensivity
follows from locality.  The constant $c$, roughly speaking, measures
the number of independent fields in the theory, at the UV fixed
point.

\item Tunneling from a meta-stable state produces a bubble, which
grows asymptotically at the speed of light, engulfing any time-like
observer\footnote{We will often use the word observer in these
lectures. We use it to mean a large quantum system with many
semi-classical observables. Quantum field theories give us models
for a host of such systems, whenever the volume is large in cutoff
units.  They are collective coordinates of large composites and have
quantum fluctuations that fall off like a power of the volume.
Quantum phase interference between different states of the
collective coordinate falls off like the exponential of the volume,
except for motions of the collective coordinates that excite only a
small number of low lying states of the system. With this definition
of the word, an observer has neither gender nor consciousness.}
propagating in the false vacuum. Inside the bubble, the state
rapidly approaches the true vacuum. If one excites a local region of
the false vacuum to sufficiently high energy, the tunneling rate
goes to infinity and meta-stability is lost. This is because the
energy density cost to produce a stable expanding bubble of true
vacuum is finite.

\item If there are two exactly degenerate quantum vacua, separated
by a barrier in field space, then, with finite cost in energy, one
can produce an arbitrarily large region of vacuum 1, in the Hilbert
space of the model which consists of local operators acting on
vacuum 2.   If the region is very large, it is meta-stable and
survives at least as long as the time it takes light to cross that
region.

\end{itemize}

\section{Are there vacuum states in models of quantum gravity?}

One of the main contentions of this lecture series is that the
answer to the above question is NO. In fact, in the end, we will
contend that each possible large distance asymptotic behavior of
space-time corresponds to a different Hamiltonian, with different
sets of underlying degrees of freedom.  This is true even if we are
talking about two different solutions of the {\it same} set of low
energy gravitational field equations. In the case of Anti-de Sitter
asymptotics we will see that the models are literally as different
from each other as two different QFTs, defined by different fixed
points.  The most conclusive evidence for this point of view comes
from the Matrix Theory\cite{bfss} and AdS/CFT\cite{malda}
formulations of non-perturbative string theory, and
ITAHO\footnote{ITAHO - In this author's humble opinion.} it is
overwhelming.  However, we can see the underlying reasons for these
differences from simple semi-classical arguments, to which this
section is devoted.

The essential point is that general relativity is not a quantum
field theory, and that the reasons for this can already be seen in
the classical dynamics of the system.  Again, it is worthwhile
making a formal list of the ways in which this is
evident\footnote{This list will use language compatible with the
idea that the quantum theory of GR is somehow the quantization of
the variables that appear in the classical Einstein equations. This
idea lies behind all attempts to define quantum gravity outside the
realm of string theory, from loop quantum gravity to dynamical
triangulations. We will argue below that this idea is wrong.}.

\begin{itemize}

\item The classical theory has no conserved stress energy tensor.
The covariant conservation law for the ``matter" stress energy is
not a conservation law, but a statement of local gauge invariance.
There is no local energy density associated with the gravitational
field. In particular, this implies that there is no gauge invariant
definition of an analog of the effective potential of
non-gravitational QFT.

\item Correspondingly, when we try to define an energy in GR, which
could play the role of the Hamiltonian in the quantum theory, we
find that we have to specify the behavior of the space-time geometry
on an infinite {\it conformal boundary}.  Geometries restricted to
such time-like or null boundaries often have asymptotic isometry
groups, and the Hamiltonian is defined to be the generator of such
an asymptotic isometry, whose associated Killing vector is time-like
or null near the boundary.  This feature of GR is the first inkling
of the {\it holographic principle}, of which much will be said
below.   It is already at this level that one begins to see that
different solutions of the same low energy effective equations will
correspond to different Hamiltonians and degrees of freedom in the
quantum theory.  I note in passing that asymptotic symmetry groups
do not seem to be an absolute necessity in this context. For
example, many of the Hamiltonians used in the AdS/CFT correspondence
have perfectly well behaved time dependent deformations and one
would suspect that these correspond to space-time geometries with no
time-like asymptotic isometries.

\item More generally, the principle of general covariance shows us
that no model of quantum gravity can have local gauge invariant
observables. This fact was discovered in string theory, and
considered an annoyance by some, long before it was shown to be a
model of quantum gravity. All known versions of string theory
incorporate this fact.  The observables are always defined on an
infinite conformal boundary.  ITAHO, the fact that other attempts to
formulate a quantum theory of gravity do not have this property, is
evidence that they are incorrect. Note that this property is in
direct contradiction with claims that a proper theory of gravity
should be {\it background independent}.  We will argue below that
the holographic principle does allow for a more local, background
independent formulation of models of quantum gravity, but that this
formulation is inherently tied to particular gauge choices.

\item More important than all of these formal properties is the
nature of the space of solutions of gravitational field theories. It
is well known that the mathematical theory of quantization begins by
identifying a symplectic structure on the space of solutions,
choosing a polarization of that symplectic structure, and
identifying a family of Hilbert spaces and Hamiltonians whose
quantum dynamics can be approximated by classical dynamics on that
phase space. The general structure of ordinary QFT is that the space
of solutions is parametrized, according to the Cauchy-Kovalevskaya
theorem, in terms of fields and canonical momenta on a fixed
space-like slice.  The corresponding formulation of GR was worked
out by Arnowitt, Deser and Misner (ADM), but it runs into a serious
obstacle.  Almost all solutions of GR are singular, and in order to
define the phase space one must decide which singular solutions are
acceptable.  There are no global theorems defining this class, but
there is a, somewhat imprecise, conjecture, called {\it Cosmic
Censorship}.  Here is what I think of as a precise formulation of
this conjecture for particular cases:

{\it Start with a Lagrangian which has a Minkowski or AdS solution
with a positive energy theorem. Consider a space-time with a
boundary in the infinite past on which it approaches Minkowski or
Anti-deSitter space, with a finite number of incoming wave packets
corresponding to freely propagating waves of any of the linearized
fluctuations around the symmetric solution\footnote{More properly,
in the Minkowski case we should probably restrict ourselves to
linearized waves that we expect to correspond to stable quantum
states in the quantum theory.}.  The amplitudes of these incoming
waves are restricted to be small enough so that the following
conjecture is true\footnote{Recall that in the quantum theory, the
classical field corresponding to a single particle has an amplitude
which formally goes to zero in the classical limit.}. The conjecture
is that to each such asymptotic past boundary condition there
corresponds a solution which obeys Cosmic Censorship: the future
evolution is non-singular, except for a finite number of finite area
black holes. The asymptotic future solution corresponds to a finite
number of outgoing wave packets plus a finite number of finite area
black holes.}

\end{itemize}

The last item focusses attention on the starring actor in the drama
that will unfold in these lectures, the black hole. Our basic
contention is that it is the answer to the age old question: How
many angels can fit on the head of a pin?  In modern language this
is phrased: {\it How many bits ($log_2$ of the number of quantum
states) can fit into a given space-time region?} This is the content
of what I will call the Strong Holographic Principle, and we will
eventually view it as a crucial part of the {\it definition} of
space-time in terms of quantum concepts.

For the moment, we stick to semi-classical arguments, and revisit
our itemized list of the properties of the QFT concept of multiple
vacua, but now with a view towards understanding whether this
concept makes sense in a theory of quantum gravity.

\begin{itemize}

\item As a consequence of general covariance, no quantum theory of
gravity can have gauge invariant correlation functions which are
localized at a point in space-time.  The physical reason for this is
the existence of black holes.  Quantum mechanics tells us that
localized measurements require us to concentrate a large amount of
energy and momentum in a small region.  General relativity tells us
that when the Schwarzschild radius corresponding to the amount of
mass (as measured by an observer at infinity) enclosed within a
sphere of radius $R$, exceeds $R$, the space-time geometry is
distorted and a black hole forms.  Bekenstein and Hawking\cite{BH}
made the remarkable observation that one can calculate the entropy
of the resulting black hole state in terms of classical properties
of the geometry.  It is given by one quarter of the area of the
horizon of the black hole, measured in Planck units. This is in
manifest contradiction with local quantum field theory, in which the
entropy scales like the volume of the sphere.  This is, in some
sense, the {\it reason} that there are no local gauge invariant
Green's functions.  The region ``inside the black hole" only has a
space-time description for a very limited proper time, as measured
by any observer in this region. We will see that a more fundamental
description is in terms of a quantum system with a finite number of
states, determined by interpreting the BH entropy as that of a
micro-canonical density matrix. The internal Hamiltonian of this
system is time dependent and sweeps out the entire Hilbert space of
states an infinite number of times as the observer time coordinate
approaches the singularity.  From the point of view of an external
observer this simply means the system thermalizes.  The external
description can be studied semi-classically and is the basis for
Hawking's famous calculation of black hole radiation. Note by the
way that Hawking radiation in asymptotically flat space-time removes
the asymmetry in our description of the classical phase space. Black
hole decay implies that once quantum mechanics is taken into account
the final states in scattering amplitudes coincide with the initial
states.

At any rate, none of the points in a local Green's function can have
a definite meaning, because we cannot isolate something near that
point without creating a black hole that envelopes the point.  It is
easy to see that the most localization we can achieve in a theory of
quantum gravity is holographic in nature.  That is, if we introduce
infinitesimal localized sources on the conformal boundary of an
infinite space-time, then straightforward perturbation theory shows
that, as long as we aim the incoming beams to miss each other
(impact parameter much larger than the Schwarzschild radius
corresponding to the center of mass energy, for each subset of
sources\footnote{Here we use language appropriate for asymptotically
flat space-time. The corresponding scattering theory for
asymptotically AdS spaces has been studied
in\cite{polchsussgidd}.}), there is a non-singular solution of the
classical field equations.  When these criteria are not satisfied,
one can prove that a trapped surface forms\cite{penroseetal}, and a
famous theorem of Hawking and Penrose guarantees that the solution
will become singular. The Cosmic Censorship conjecture implies that
this singularity is a black hole, with a horizon area bounded from
below by that of the trapped surface.

In quantum field theory, the regime of scattering in which all
kinematic invariants are large, is dominated by the UV fixed point.
In this regime the differences between different vacuum states
disappear.  In quantum gravity by contrast, this is the regime in
which black holes are formed. In asymptotically flat space, the
specific heat of a black hole is negative, which means that at
asymptotically high energies, the black hole temperature is very
low. Thus, the spectrum of particles produced in black hole
production and decay depends crucially on the infrared properties of
the system.   Different values of the {\it moduli}, the continuous
parameters that characterize all known asymptotically flat string
theory models, correspond to different low energy spectra. So in
theories of quantum gravity, scattering at large kinematic
invariants depends on what some would like to call the vacuum state.
This is our first indication that these parameters correspond to
different models, not different quantum states of the same system.

Black holes also falsify the claim that the high temperature
behavior of the partition function is dominated by a conformal fixed
point.  In fact, all conformal field theories have positive specific
heat and a well defined canonical ensemble.  The negative specific
heat of black holes in asymptotically flat space-time implies that
their entropy grows too rapidly with the energy for the canonical
partition function to exist.  Although black holes are unstable,
they decay by Hawking radiation, and the Hawking temperature goes to
zero as the mass of the hole goes to infinity.  Thus the high energy
behavior of the micro-canonical partition function in asymptotically
flat space would appear to be dominated by black holes, and cannot
be that of a CFT.

It is interesting to carry out the corresponding black hole entropy
calculation in the other two maximally symmetric space-times, with
positive or negative values of the c.c. .  The modified
Schwarzschild metric is

$$ds^2 = - (1 - V_N (r) \pm (\frac{r}{R})^2 ) dt^2 + \frac{dr^2}{(1 - V_N (r) \pm (\frac{r}{R})^2
)} + r^2 d\Omega^2 ,$$ where $V_N (r)$ is the Newtonian potential in
$d$ space-time dimensions,
$$ V_N (r) = c_d \frac{G_N M}{r^{d-3}},$$ $R$ the radius of
curvature of the de Sitter or AdS space, and the $+$ sign is for the
AdS case. In that case, the horizon radius is the unique zero of
$g_{tt}$.  When it is much larger than $R$ it is given approximately
by
$$R_S^{d - 1} \sim c_d MR^2 L_P^{d-2} , $$ where the Planck length
is defined by $G_N = L_P^{d-2}$, in units where $\hbar = c = 1$. The
area of the horizon is $A_d R_S^{d-2} $, so the BH entropy is

$$ B_d (MR)^{\frac{d-2}{d-1}}(\frac{R}{L_P})^{\frac{d-2}{d-1}} .$$
$B_d = A_d c_d^{\frac{d-2}{d-1}}$.  Remarkably, this looks like the
entropy formula for a conformal field theory in ${\bf d-1}$
dimensions, living on a space with volume $ \sim R^{d-2} $.  In this
interpretation, the quantity $(\frac{R}{L_P})^{\frac{d-2}{d-1}}$
plays the role of ``the number of independent fields" in the CFT.
This formula is one of the key elements of the AdS/CFT
correspondence \cite{malda}\cite{gkpw}\cite{aharony}.  Note in
particular {\it the dependence of the high energy density of states
on the c.c.}.  In bulk QFT, which motivates the idea of different
vacuum states, the c.c. is a low energy property of the theory and
the high energy density of states does not depend on it.  We will
see that the manifold examples of the AdS/CFT correspondence make it
abundantly clear that different solutions of the bulk field
equations correspond to different quantum Hamiltonians; different
models of quantum gravity rather than different states in a given
model.

The dS case is even more striking.  $g_{tt}$ has two zeroes, the
larger of which is the cosmological horizon, which persists even
when the black hole mass goes to zero. The sum of the areas of those
two horizons is always less than that of the cosmological horizon of
``empty dS space", and in fact decreases as the black hole mass
increases.  There is a maximal mass (Nariai) black hole, whose two
horizons have equal area. When combined with the result of Gibbons
and Hawking\cite{GH}, that the dS vacuum state is a thermal state
for the local observer in a maximal causal diamond of dS space, this
result leads to the conclusion\cite{tbwf} that a quantum theory of a
stable dS space must have only a finite number of quantum states.

\item The semi-classical theory of quantum tunneling in the presence
of gravity begins with the seminal paper of Coleman and De
Lucia\cite{cdl}. It confirms the picture of different solutions
corresponding to different models, rather than different states,
although almost all of the literature is couched in the language of
vacuum decay. I will use the terms {\it true and false minima}
rather than {\it true or false vacua} in order to emphasize that the
conventional interpretation is wrong. The characteristics of
gravitational tunneling depend crucially on the values of the energy
density at the true and false minima.  Let us begin with the case
where the true minimum has negative c.c. .  One of the most
important results in \cite{cdl} is that in this case, the classical
evolution after tunneling does not settle down to the AdS solution
with the field sitting at the true minimum.  Instead, the geometry
undergoes a singular Big Crunch. There is no conserved energy, and
as the universe inside the bubble contracts, the energy of the
scalar field gets larger and larger.  The field explores its entire
potential and does not remain near the ``true minimum".  More
importantly, the semi-classical approximation breaks down.  Even in
quantum field theory, particle production occurs and one might
imagine that fluctuations in the energy density could lead to black
hole formation.  We will reserve to a later section a conjecture
about what the real physics of the singularity is.  For now we only
note that the maximal causal diamond in this crunching geometry has
only finite proper time between its past and future tips, as well as
a maximal finite area for any space-like $d - 2$ surface on its
boundary.

The main point here is that there is no sense in which this
semi-classical approximation describes decay to a well understood
ground state.  Below, by using the holographic principle, we will
find a sensible interpretation of some of these processes (but not
as decays) and present arguments that others simply can't occur in
well defined models of quantum gravity. This is in stark contrast to
the situation in QFT, where of course the value of the potential at
its minimum is unobservable.  Notice that none of this has anything
to do with the AdS solution, which one gets by fixing the scalar
field at its true minimum.  This solution may or may not represent a
sensible model of quantum gravity, but it certainly has no
connection to the hypothetical model in which the CDL instanton
describes some kind of transition.

When the true minimum has positive c.c., the situation is much
better. Classical evolution of the scalar field after tunneling,
rapidly brings it to rest at the true minimum.  Furthermore, the
resulting space-time has an (observer dependent) cosmological
horizon.  Inside an observer's horizon volume, all fields rapidly
approach the empty dS configuration.  We will see below that in this
case of dS to dS tunneling, more can be gleaned from the nature of
the semi-classical CDL solution, and it is all consistent with the
idea that the quantum theory of stable dS space has a finite number
of of states.

The case of a true minimum with vanishing c.c., whether this is
achieved at a finite point in field space, or at asymptotically
infinite scalar field, is much more ambiguous. If the falloff of the
potential is that found in all asymptotic regions of string theory
moduli space\footnote{As we will emphasize below, the notion of a
potential on string theory moduli space is a problematic one.
Nonetheless, if one accepts the validity of the concept one can use
the symmetries of string theory to establish bounds on the behavior
of the potential at infinity\cite{bdcosmoetal}.  } then the future
causal boundary of the universe is similar to that of Minkowski
space: the maximal causal diamond has infinite area holographic
screen, and at finite points within that diamond, at late times, the
space-time curvature goes to zero, and the scalar field asymptotes
to the zero c.c. point.  On the other hand, this is NOT an
asymptotically flat space.  Furthermore, if one takes the analogues
of outgoing scattering states for this universe, then most do not
extrapolate back to smooth perturbations of the instanton geometry.
The meaning of this kind of situation is the central issue in trying
to establish the existence of the String Landscape.  We will explore
these issues, which are far from settled, in section 4 below.

To summarize, CDL tunneling provides abundant evidence for the fact
that AdS solutions of gravitational field equations are NOT part of
the same model as other stationary points of the same effective
action. {\it One never tunnels to AdS space.} It also suggests that
there can be models of quantum gravity with a finite number of
states, which describe stable dS space. We will complete that
discussion in section 6. Similarly, there is no tunneling to
asymptotically flat solutions of the field equations, which again
must be regarded as (possibly) defining separate models of quantum
gravity.  We will argue below that there is no tunneling {\it from}
AdS minima either, and that tunneling from asymptotically flat
minima leads to a bizarre picture of the final state.

\item Finally, we revisit the question of creating large meta-stable regions of
space, which are in ``another vacuum".  If we start from an
asympotically flat, or AdS minimum, and the potential is everywhere
much less than the Planck scale and varies on a field space scale
$\leq m_P $, then it is easy to find finite energy incoming
configurations which move the field into another minimum over a
sphere of radius $R$. However, if there is any potential barrier at
all between the asymptotic minimum of the potential and the field
value inside $R$, then the domain wall energy will scale like
$R^{d-2}$ and the Schwarzschild radius of the configuration will be
$> R$.  In other words, a black hole will form before the false
vacuum bubble gets too big.  Notice that if the false vacuum is a dS
space, there will be an additional, volume contribution to the
Schwarzschild radius.  This guarantees that the black hole ALWAYS
forms before the bubble can inflate\footnote{A lot of confusion is
caused by solutions of the equations of GR which describe an
arbitrarily large region of dS space or slow roll inflation,
connected to an asymptotically flat or AdS region with a small black
hole in it. These solutions cannot evolve from data that is
non-singular in the past and in particular from incoming scattering
data in a space-time with a well defined past conformal infinity. If
they represent anything in a real quantum theory of gravity it is an
artificially entangled state of two, generally different, quantum
Hilbert spaces.  In the present discussion we approach localized
regions of false minimum by starting from small regions that do not
inflate and boosting the incoming energy continuously.  In that case
the black hole mass is bounded from below by the integrated energy
density of the false minimum and the black hole always forms before
inflation can occur.} Thus, while auxiliary minima of a
sub-Planckian effective potential do allow the creation of
meta-stable states, they are not false vacua. The meta-stable
regions that resemble homogeneous vacuum solutions are of limited
size.  Anything above that size is a black hole, which is to say, a
thermodynamic equilbrium state indistinguishable from any other
state of the theory that maximizes the entropy within the region
$\leq R_S$. Notice also that there is no sense in which the decay of
the meta-stable states created here is related to the instanton
transitions discussed above. These are localized excitations of the
true vacuum state, and will decay back to it by radiating particles
off to infinity.

\end{itemize}

The conclusion is that rather simple classical considerations show
that, whatever the theory of quantum gravity is, {\it it is not a
QFT and the QFT concept of a vacuum state does NOT generalize to QG.
Different solutions of the same low energy effective gravitational
field equations, can correspond to different models of QG, rather
than different states of the same model.}

\section{Matrix Theory and the AdS/CFT correspondence}

Indeed, all of our non-perturbative constructions of quantum gravity
have this property in spades.  In this section I'll quickly review
these constructions, starting with the case of asymptotically flat
space.

\subsection{Matrix theory}

We have seen that in $d$ dimensional asymptotically flat space, the
entropy grows like $E^{\frac{d-2}{d-3}}$, so that conventional
constructions of the partition function and the path integral fail.
However, at least if $d > 4$, the light-front partition function at
fixed longitudinal momentum

$${\rm Tr} e^{ - \beta P^- } ,$$
should be well defined, and we might hope to discover a more or less
standard Lagrangian formulation of gravity.  The Lagrangian for a
single supersymmetric particle (superparticle) in 11 dimensions, is

$$\int dt\ \frac{p}{2} \dot{\bf x}^2 + i \theta \dot{\theta} .$$
Here  $p \geq 0$ is the longitudinal momentum, which is treated as a
fixed constant, and the time variable is light front time. ${\bf x}$
is a transverse 9-vector and $\theta_a $ a $16$ component light
front spinor. The system is quantized in terms of $9$ commuting
transverse momentum variables ${\bf p}$ and the $16\ \theta_a$, with
commutation relations

$$ [\theta_a , \theta_b ]_+ = \delta_{ab}.$$ The SUSY generators and
Hamiltonian are $$q_a = \theta_a ,$$ $$Q_a = ({\bf \gamma \cdot
p})_{ab} \theta_b ,$$ $$P^- = \frac{({\bf p})^2}{2 p} .$$ The
$\theta_a$ don't appear in the Hamiltonian, which describes a single
massless relativistic particle.  However, they give this massless
state a degeneracy, with precisely the spin content of the 11
dimensional SUGRA multiplet.

Notice that this procedure only makes sense when $p$ is strictly
greater than zero. Particles with zero longitudinal momentum are
non-dynamical.  However, when the longitudinal momentum is
continuous, the region of low longitudinal momentum becomes singular
and one must exercise great care in treating it in order to extract
correct results. In QFT this is often done by the method of Discrete
Light Cone Quantization (DLCQ), in which the longitudinal direction
is formally compactified so that $p$ takes on only the discrete
values $\frac{n}{R}$, with $n$ a positive integer. One then studies
the limit $R \rightarrow \infty$, by considering wave packets made
from states including large values of $n$, so that they are
localized in the longitudinal direction and can become independent
of $R$.  One convenience of this procedure, often exploited in QFT
is that for fixed total longitudinal momentum $\frac{N}{R}$, a
multi-particle state can have only a finite number of particles in
it, so that in DLCQ field theory is approximated by the quantum
mechanics of a finite number of particles.

The word approximated in the previous paragraph has to be stressed.
The real system is obtained only in the limit when $N$ is strictly
infinite. Thinking about multi-particle states, we see in particular
that, at fixed $R$, only those states with light cone energies $\sim
\frac{1}{N}$ will survive in the limit $N \rightarrow\infty , R
\rightarrow \infty $, with $p = \frac{N}{R}$ fixed. This introduces
a degree of ambiguity into DLCQ, which can be exploited to simplify
the limit.  This ambiguity, well known in QFT, has mostly been
ignored in the gravitational case, because Seiberg, following work
of Sen and Susskind \cite{sss} found a particularly compelling form
of DLCQ using string dualities.

Matrix Theory proceeds from this kinematical framework, by
introducing an alternative to the Fock space treatment of identical
particles. Instead, we generalize the variables ${\bf x}$ and
$\theta_a $  to simultaneously diagonalizable $N \times N$ matrices.
These can be written as
$${\bf X}  = \sum {\bf X}^I e_I , \quad\quad\quad \Theta_a = \sum
\theta_a^I e_I , $$ where $e_I^2 = e_I$ and ${\rm Tr} e_I = n_I$,
$\sum n_I = N$. This representation is redundant if some of the
$n_I$ are the same, and we have a gauge symmetry permuting the $e_I$
with equal trace, which is precisely the gauge symmetry of particle
statistics. The fact that half integral spin is carried by the
anti-commuting variables $\Theta_a$ guarantees that the
spin-statistics connection is the conventional one.  The Lagrangian
is

$$L = \frac{1}{R} {\rm Tr}\ \bigl{[} \frac{1}{2} \dot{\bf X}^2 + i
\Theta \dot{\Theta} \bigr{]},$$ and as we run over all possible
choices of the $e_I$ we reproduce the Lagrangians for $k \leq N$
supergravitons with all configurations allowed in DLCQ, and total
momentum $\frac{N}{R}$.

If we insist on preserving all the SUSY, as well as the $SO(9)$
symmetry of this Lagrangian, there is a unique way of modifying it
that allows for interaction between the supergravitons. To see what
it is, we note that the Lagrangian we've written down is the
dimensional reduction of $N=4$ super Yang-Mills theory with gauge
group $U(1)^N \ltimes S_N$, which is the low energy effective
Lagrangian on the maximally Higgsed Coulomb branch moduli space of
$U(N)$ SYM theory. The Lagrangian is written in temporal gauge (with
the time of the gauge theory identified with light cone time) and
the restriction to Bose or Fermi statistics for the particles is
just the residuum of the Gauss Law of the non-abelian gauge theory.
The full non-abelian Lagrangian is

$$L = \frac{1}{R} {\rm Tr}\ \bigl{(} \frac{1}{2} (D_t {\bf X})^2
- g^2 [X^i , X^j ]^2 + i\Theta D_t \Theta + g [{\bf \gamma \cdot
 X},{\Theta}] \bigr{)},$$ where the adjoint covariant derivatives
 are $D_0 Y = \partial_t Y + g [A_t , Y].$  The constraints are now
 obtained by varying w.r.t. $A_t$ and then setting $A_t = 0$. We'll
 describe how the SYM coupling is determined in terms of the Planck
 length below.

 Before doing so, we note that this Lagrangian can also be shown to
 be the world volume Lagrangian of D0 branes in ten dimensional Type
 IIA string theory.  The excitations on D-branes are open strings
 satisfying the appropriate mixed Dirichlet/Neumann boundary
 conditions.  For N D-branes, the lowest excitations in open superstring theory have
 the quantum numbers of the maximally supersymmetric $U(N)$
 Yang-Mills multiplet. If all of the spatial boundary conditions are
 Dirichlet, then the low energy world volume Lagrangian is unique
 and is given by the above formula.   This idea led Seiberg,
 following Susskind and Sen, to argue that the compactified theory
 was just given by the D0 brane Lagrangian on the compact space.
 This conjecture is valid if we preserve at least 16 supercharges.
 It identifies the correct degrees of freedom, and their Lagrangian
 is completely determined by symmetries.

 The D0 brane picture tells us how to identify the Yang-Mills
 coupling.  Interactions are determined by the open string coupling,
 $g_S$, so $g_{YM}^2 = g_S$.  On the other hand, Type IIA string
 theory is the compactification of M-theory (the quantum theory
 whose low energy limit is 11D SUGRA), on a circle whose radius is
 small in Planck units.  D0 brane charge is Kaluza-Klein momentum.
 So we have the identification

 $$ \frac{l}{l_S g_S} \propto \frac{1}{R} ,$$ where we've equated
 the string theory formula for the D0 brane mass to the KK formula.
In the duality between M-theory and Type IIA string theory, the
string is viewed as an M2 brane wrapped on the small circle, so
$$l_S^{-2}  \propto l_P^{-3} R .$$  Combining the two formulae we
find $g_S \propto (R / l_P )^{3/2} .$ All of these formulae are
actually exact consequences of SUSY, so the constants we have
omitted can be calculated exactly.

Seiberg's prescription tells us that if we want to find the DLCQ of
M-theory compactified on a torus or K3 manifold, we should study the
world volume Lagrangian of D0 branes moving on that manifold.  If
the manifold has size of order the 11D Planck scale, then it is very
small in string units, and we should do a T-duality transformation
to find a description that is under greater control.   For a torus
of less than four dimensions, this gives us SYM theory compactified
on the dual torus. These are all finite theories and the
prescription is unambiguous.   Many exact results, including some
famous string dualities can be derived from this prescription, and
agree with calculations or conjectures that one already had in
supergravity or string theory.  Other calculations, not protected by
supersymmetry non-renormalization theorems are only supposed to be
correct when takes the $N \rightarrow \infty$ limit, keeping only
states whose light cone energy scales like $\frac{1}{N}$.

For a four torus or a K3 manifold, one naively gets the four
dimensional SYM theory, which is not renormalizable.   However, the
T-dual string coupling is large, so we should really be studying the
D4 branes (into which the D0 branes are converted by T-duality) in
the strong coupling limit.  In this limit, D4 branes become M5
branes.  The world volume theory on N M5 branes is a maximally
superconformally invariant $6$ dimensional theory.  It is
compactified on $T^5$ or $K3 \times S^1$.  Again, the prescription
is finite and makes a number of correct exact predictions.  It is
however more difficult to calculate with since not much is known
about the $(2,0)$ superconformal field theory.

If we add one more circle to either of these constructions, we
obtain {\it little string theory}.  This is the world volume theory
of $N$ NS5 branes in the zero string coupling limit. Even less is
known about this model than about the $(2,0)$ superconformal field
theory, and there have even been questions raised about whether it
really exists.   With six or more compact dimensions, the Seiberg
construction fails and we do not have a working definition of the
DLCQ of M-theory with $4$ or $5$ asymptotically flat dimensions.

Among the most striking features of these constructions is that each
different gravitational background gives rise to a different quantum
Hamiltonian.   Even two versions of M-theory with values of
continuous moduli that differ by a finite amount, correspond to the
same field theory Hamiltonian on different compactification
manifolds.  And remember that the canonical variables of this
Hamiltonian do not include a gravitational field.  The geometry of
the compactification manifold is not a dynamical variable in the
Matrix theory Hamiltonian.

\subsection{The AdS/CFT correspondence}

The correct statement of the AdS/CFT correspondence is that in
certain quantum field theories in $d-1$ space-time dimensions, there
is a regime of large parameters, in which three important properties
are satisfied:

\begin{itemize}

\item The high temperature behavior of the partition function on a
spatial sphere of radius $R$ is $ c (RT)^{d-2}$, with $c \gg 1$.

\item The dimension of most operators at the UV fixed point which
defines the theory go to infinity.

\item The Green's functions of those operators whose dimension
remains finite can be computed approximately by solving the
classical field equations of a $d + D$ dimensional gravitational
Lagrangian, with boundary conditions first outlined by \cite{gkpw}.
The space-time metric has a conformal boundary identical to that of
$AdS_{d} \times K$, where $K$ is a compact manifold.  If the
non-compact space-time is exactly $AdS_d$ then the boundary field
theory is conformal.

\end{itemize}

As a consequence of the last property, we consider such QFTs to be
{\it definitions} of models of quantum gravity, with fixed
asymptotic background. The idea that AdS/CFT defines a duality
between two independently defined theories, is probably without
merit. For a subclass of these theories, one of the large parameters
is an integer $N$ which controls the size of the the gauge group of
the boundary field theory, and the model has a conventional large
$N$ expansion.  In this case there is a weak coupling string theory
description of the model, which goes beyond the classical gravity
expansion described above. In these cases, the models have at least
two adjustable parameters.  One, $N$, controls the standard planar
expansion of the theory, which can be recast as an expansion in
world sheet topology. The other, loosely called the {\it 't Hooft
coupling}, is continuous (at least in the large $N$ limit).  When it
is large, the solution of the theory in terms of classical
gravitational equations is valid.  When the 't Hooft coupling takes
on moderate or small values there is a calculation of the
correlation functions of all operators whose dimensions are finite
in the large $N$ limit, in terms of a world-sheet quantum field
theory. In most of the interesting cases\footnote{The models of
\cite{natietal} are an exception.} the world sheet theory is hard to
solve, but enormous progress has been made in establishing the
conjecture.

However, even if we were able to calculate everything, including all
higher genus contributions in the world sheet theory, this would not
constitute an independent definition of the ``other side" of the
``AdS/CFT duality".  String perturbation theory is a non-convergent
asymptotic expansion.  We know plenty of examples where its
existence and finiteness to all orders is {\it not} a guarantee of
the existence of a real quantum model of gravity. Bosonic matrix
models related to $1 + 1$ dimensional string theories are a
calculable example\cite{oldmatrix}. A continuous infinity of other
examples is provided by moduli spaces of $4$ dimensional
compactifications of string theory with $N=1$ SUSY. These have well
defined perturbation expansions. However, general symmetry
arguments, as well as many explicit instanton calculations show that
there must be a non-perturbative superpotential on this moduli
space, if this set of models makes sense at all. This means that all
of the perturbation expansions except {\it perhaps} for a discrete
set of points in moduli space do not correspond to well defined
models. Furthermore, even if the wildest conjectures about the
string theory Landscape are correct, most of these discrete points
correspond to space-times with non-zero c.c..  This means that the
flat space S-matrix elements one calculates in string perturbation
theory do not correspond, even qualitatively, to the correct set of
observables of the hypothetical underlying model.  We will return to
this point when we discuss the string Landscape below.   Our
conclusion here is that the AdS/CFT correspondence is a {\it
definition} of a class of models of QG, in terms of QFTs defined on
the conformal boundary of AdS space.

It is important to emphasize that most QFTs fit into neither of
these categories, even when they have a large $N$ expansion.  All
large $N$ models, and many other examples, such as the tensor
product of any large collection of mutually non-interacting QFTs (or
theories that are small perturbations of such a collection) satisfy
the first of our criteria above. Referring back to the formula for
black hole entropy in AdS space, we see that this criterion can be
rephrased as: {\it AdS/CFT gives a rigorous justification of the BH
entropy formula for asymptotically AdS space-times}. Comparison of
the two formulae leads to the conclusion that the constant $c$ is a
measure of the ratio of the AdS curvature radius to the Planck
length. Obviously, any classical space-time interpretation of the
model will be valid only when this parameter is large, but this is
only a necessary condition for the classical gravity approximation
to be valid.

To understand better what is going on, let's recall the basic
equations of the AdS/CFT correspondence.  The Euclidean\footnote{The
Euclidean rotation familiar from QFT is not valid for QG in
asymptotically flat space, because the density of states blows up
too rapidly for the finite temperature partition function to be well
defined.  However, in AdS space the quantum theory is a boundary QFT
and the Wick rotation makes sense.} AdS metric is
$$ ds^2 = (1 + \frac{r^2}{R^2}) d\tau^2 + \frac{dr^2}{1 +
\frac{r^2}{R^2}} + r^2 d\Omega_{d-2}^2 . $$

It follows that, at large $r$, solutions of the Klein-Gordon
equation behave like $r^{\lambda_{\pm}} J(\tau, \Omega), $ with
$$\lambda (\lambda + d - 1) = m^2 R^2 .$$  The $\pm$ signs refer to
the two roots of these equations.  The AdS/CFT prescription is to
solve the coupled non-linear Einstein matter equations, with the
boundary conditions that the fields behave like the larger root of
this equation, and arbitrary source function $J$.  Analogous
boundary conditions are imposed on the metric and other higher spin
fields. The action as a functional of the source is the generating
functional for conformally covariant Green's functions on the
boundary.

A consequence of this prescription is that every primary operator in
the boundary CFT corresponds to a different field in the bulk.  The
mass of small fluctuations is related to the dimension of the
primary.  Thus, the bulk theory will have, generically, an infinite
number of fields. The only known way to write an approximately local
field theory with an infinite number of fields in AdS space, is to
consider field theory with a finite number of fields on $AdS \times
K$, where $K$ is a compact manifold. The infinity then corresponds
to a complete set of functions on $K$. The degeneracy of the
Laplacian on $K$ for high eigenvalues is power law in the
eigenvalue, so this prescription could at most give us a power law
growth of the number of fields of mass $m$, as $m\rightarrow\infty$.

It is well known that the number of primary operators of dimension
$D$ grows exponentially with a power of dimension, which implies an
exponentially growing number of {\it fields}, in the approximate
local field theory describing fluctuations around the hypothetical
$AdS \times K$ background. Kaluza-Klein compactification on $K$
gives rise only to a spectrum of masses that grows like a power of
the mass (in $1/R$ units, where $R$ is the radius of curvature of
$AdS$, typically of the same order of magnitude as that of $K$).  In
examples where the CFT is dual to a weakly coupled string theory,
such an exponential growth {\it is} seen among string states. So,
for a generic CFT, one needs parametrically large entropy in order
to claim that the geometrical radii are larger than the Planck
length, but also another large parameter to guarantee that
geometrical radii are larger than the length defined by the string
tension.

It should be emphasized that very few CFTs actually correspond to
weakly coupled string theories. The necessary and sufficient
condition is that the theory have a conventional {\it matrix} $1/N$
expansion.  This is what is necessary to have both a free string
limit, and a topological structure of interactions that corresponds
to a sum over world sheet topologies.  Neither vector large N
limits, nor the topological expansions typical of theories with
comparable numbers of flavors and colors, or matter in other large
representations of $SU(N)$\footnote{The second rank symmetric and
anti-symmetric tensor representations of $O(N)$, {\it do} appear in
orientifold projections of large $N$ gauge theories and have a
string loop expansion.}, have a free string interpretation. Thus,
for many CFTs, there seems to be no interpretation of their
correlation functions as a set of observables corresponding to
objects propagating in an AdS space\footnote{Even when the CFT has
an entropy and dimension spectrum corresponding to an AdS radius
that is large compared to both the ``string length" and the Planck
length, in the sense described above, it may not have a simple space
time interpretation. A simple example is maximally supersymmetric
$SU(N) \times SU(M)$ Yang Mills theory with both 't Hooft couplings
large, or a perturbation of it by an exactly marginal operator
constructed as a product of relevant operators from the individual
theories.}.

In all rigorously established examples of the AdS/CFT correspondence
the large parameter is an analog of the 't Hooft coupling of a large
$N$ gauge theory, a parameter which is continuous in the planar
limit.  In the two and three dimensional examples the 't Hooft
coupling is really a ratio of two large integers, while in four
dimensions it is the rescaled Yang Mills coupling.  It is important
that the theory is conformally invariant for {\it every} value of
the 't Hooft coupling.  In the limit when the coupling is large some
dimensions remain of order $1$, while others go to infinity.
Furthermore, the multiplicity of operators with order $1$ dimension
grows only like a power of the dimension, consistent with a bulk
space-time interpretation on a background of the form $AdS \times
K$.   All of the examples where this behavior has been established
are exactly supersymmetric.

Non-supersymmetric marginal perturbations of these theories {\it
all} lead to models with at most isolated fixed points at 't Hooft
coupling of order one.  One can also consider orbifolds of the
${\cal N} = 4$ SYM theory, whose planar diagrams coincide with the
original theory, and are conformally invariant for all values of the
't Hooft coupling.  However, the leading non-planar corrections to
the beta functions of several couplings are non-zero and depend
explicitly on the 't Hooft coupling.  The theories will be
conformal, if at all, only at 't Hooft coupling of order one. These
theories provide interesting analogs of tachyon free
non-supersymmetric string theories in flat space-time.  Those
asymptotically flat models seem completely sensible at string tree
level, but the loop diagrams are divergent.  If one tries to invoke
the Fischler-Susskind mechanism to cancel these divergences, one
finds perturbations of the space-time geometry which are singular in
either the remote past or future or both.  The string perturbation
expansion breaks down.   There is no evidence that these models
really exist. The same is true for the non-supersymmetric orbifold
theories. At leading order in the planar expansion, we have a free
string theory on an AdS space-time. Finite string coupling
corrections destroy this interpretation, except perhaps for a
particular AdS radius of order the string scale. The question of
whether the model at this particular radius makes sense is the
question of whether the leading non-planar beta function has a
finite coupling fixed point. In fact, that only guarantees that
string perturbation theory in AdS space will make sense at that
radius, and one must confront the resummation of the divergent
$\frac{1}{N}$ expansion.

\subsection{Domain walls and holographic renormalization group flow}

When a flat space QFT has two isolated degenerate vacua,
$\phi^i_{\pm}$ it also has domain wall solutions in which the scalar
fields vary only in a single coordinate $\phi^i (z)$, and $\phi^i
(\pm\infty) = \phi^i_{\pm} $.  These solutions are stable and have a
finite surface energy density, called the tension of the domain
wall. They are limits of meta-stable finite energy states of the
field theory with spherical domains of one vacuum inside the other.
We have already argued that no such limit exists in theories with
gravity.  If the spherical domain wall becomes too big it collapses
into a black hole.

There are however many examples of infinite hyper-planar domain wall
solutions of Lagrangians with gravity, and the AdS/CFT
correspondence gives us a novel interpretation of them. Consider a
scalar field coupled to gravity with a potential having two
stationary points, one a maximum and one a minimum, both with
negative c.c. .   There are AdS solutions corresponding to each of
these points, and it is possible for both of them to be stable.
Indeed Breitlohner and Freedman showed that tachyonic scalar fields
are allowed in $AdS_d$ space, as long as the tachyonic mass
satisfies
    $$4 m^2 R^2 > (d - 1)^2 .$$ Referring to the dictionary relating
bulk masses to boundary dimensions, we see that this is the
condition for dimensions to be real and that B-F allowed tachyons
are dual to {\it relevant} operators.

The equations for a domain wall solution connecting the two
stationary points are
$$\phi^{\prime\prime} (z) + (d -
1)\frac{\rho^{\prime}}{\rho}\phi^{\prime} (z) = \frac{dV}{d\phi} .$$
$$2\rho^{\prime\ 2} = \epsilon^2 \rho^2 (\phi^{\prime\ 2} - V) .$$
We have rescaled fields and coordinates so that everything is
dimensionless (see the discussion of instanton solutions below) and
$\epsilon$ is a measure of the rate of variation of the potential
$V$ in Planck units.  The metric is
$$ds^2 = dz^2 + \rho^2 (z) d{\bf x}^2 ,$$ corresponding to a
hyper-planar domain wall geometry.  Our boundary conditions are that
$\phi (\pm \infty)$ be the positions of the two stationary points.
$\rho$ then interpolates between the two different $AdS$ geometries.

Near the AdS maximum of the potential, the two solutions of the
linearized equation both fall off at infinity, so we only use up one
boundary condition by insisting that the solution approach the
maximum as $z \rightarrow \infty $.  The solution then contains both
possible power law behaviors and thus, {\it from the point of view
of the AdS/CFT correspondence at the maximum, it corresponds to a
perturbation of the Lagrangian of the boundary field theory by a
relevant operator.}  This is a novelty compared to traditional QFT.
The domain wall is not an infinite energy state in the original
model, but a perturbation of its Hamiltonian.   It becomes clear
that we should be trying to view the domain wall solution as the
``anti-holographic" representation of a boundary renormalization
group flow between two CFTs.

In general there is no such solution.   The problem is that one of
the linearized solutions of the fluctuation equations around the AdS
minimum blows up at infinity.  Thus we need two boundary conditions
to ensure that the solution approaches the minimum as $z
\rightarrow\infty$ {\it and} that its derivative goes to zero there.
Having used up one parameter on the other side of the wall, we do
not have this freedom. However, we can always find a solution by
fine tuning one parameter in the potential, in order to set the
coefficient of the growing mode to zero.  Thus, the space of
potentials with static domain wall solutions connecting two AdS
stationary points is co-dimension one in the space of all potentials
with two such stationary points\footnote{For future reference, we
note that the parameter counting remains the same when we search for
a domain wall connecting an asymptotically flat minimum to one with
negative c.c., although the interpretation of the solution as an RG
flow is no longer applicable.}.   Note that the fact that the second
stationary point is a minimum is consistent with, and implied by,
the RG interpretation.  An RG flow should always approach its IR
fixed point along an irrelevant direction in the space of
perturbations of that fixed point.  The AdS/CFT dictionary
translates {\it irrelevant} as {\it positive mass squared}.

Having found such an RG flow we are almost ready to declare that we
have a self consistent discovery of a new CFT with a large radius
AdS dual.  However, consistency requires that we check all
directions in the bulk scalar field space, to determine if there are
any tachyonic modes that violate the B-F bound.  One way to
guarantee both the existence of the domain wall solution and its B-F
stability is to work in SUGRA, and insist that both stationary
points preserve some SUSY.  A host of solutions of this type have
been found, that interpolate between fixed points with different
numbers of supercharges in their super-conformal algebra.

Remarkably, when we perturb a supersymmetric CFT with a large radius
dual by a relevant operator that violates all supersymmetry, we have
yet to find a consistent solution.  There are a number of smooth
domain wall solutions of this type, but one always finds tachyons
that violate the B-F bound in the spectrum of scalar fluctuations of
the new minimum.  There is, as yet, no theorem that this is {\it
always} the case, but when combined with the failure to find
non-supersymmetric large radius AdS spaces by orbifolding one is led
to suspect a connection between SUSY and the low curvature of
space-time.  We will see below that the construction of holographic
space-time seems to imply that all consistent theories of gravity in
asymptotically flat space are exactly supersymmetric.  Many years of
failure to find consistent perturbative string constructions, which
violate SUSY in asymptotically flat space-time, have convinced most
theorists that no such theories exist.

By contrast, the theory of the String Landscape, to which we will
turn in a moment, suggests that there is no particular relation
between the size of the cosmological constant and the scale of SUSY
breaking. This effective field theory based scenario seeks to
identify a huge set of string models with many independent small
positive contributions to the effective potential.  Adding these to
a large negative contribution, one argues that if the number of
positive contributions is of order $10^ X$ with $ X$ significantly
larger than $123$, then there will be many of these models with
positive c.c. of order the one we observe. One then invokes the
``successful" anthropic prediction of the c.c. to explain why we
happen to see only this special class of models. As a byproduct,
this construction produces a huge set of models with very small
negative c.c., without SUSY.   Indeed the typical strategy is to
find such a negative c.c. AdS solution and then add a single small
positive contribution to get a model representing the real world.

It thus seems rather important to determine whether there are in
fact non-supersymmetric CFTs with large radius AdS duals.  This is a
well defined mathematical problem, in stark contrast to the
effective potential discussion of the landscape.  It's my opinion
that more people should be working on it.

\section{Is there a string theory landscape?}

The basic idea of the string landscape is easy to state.   If one
looks at compactifications of string theory to four dimensions, with
$N \geq 2$ SUSY, we find moduli spaces of models of quantum gravity,
with continuous parameters.  The number of such parameters is
related to the topological complexity of the compactification
manifold.  For example, in compactifications of Type IIA string
theories on Calabi-Yau manifolds, we find a vector multiplet of N=2
SUSY for every non-trivial $(1,1)$ cycle\footnote{Actually, it's a
bit more complicated.  One of the vector fields is part of the N=2
SUGRA multiplet.   We get a number of non-gravitational vector
multiplets equal to $h_{1,1} - 1$.}  We find a massless
hypermultiplet for every $(2,1)$ cycle.  So complicated topologies
have high dimensional moduli spaces.

When we consider compactifications with only N=1 SUSY, for example
heterotic strings on $CY_3$, then we find a similar list of moduli
at string tree level and to all orders in perturbation theory.
However, there is no non-perturbative argument (in most cases) that
these moduli spaces are an exact property of the theory.  The fact
that there are moduli spaces in perturbation theory is related to a
continuous shift symmetry of the superpartner of the dilaton field.
There are many non-perturbative effects that violate this symmetry.
Thinking in terms of low energy effective field theory, we imagine a
non-trivial superpotential on this moduli space, which leads to a
non-trivial potential.   A generic function on a space of dimension
$D$ is expected to have a set of local minima whose number is
exponential in $D$.  This is the most naive picture of the string
landscape.

As we have described it, the landscape is not under any quantitative
control.   If one tries to write down a few terms in the expansion
of the superpotential around weak coupling, one finds that
non-trivial minima always lie at values of the string coupling where
the expansion is invalid.  A major step in the development of
landscape ideas was the notion of {\it flux
compactification}\cite{fluxcomp}.  This was the study of solutions
in which field strengths of p-form fields on the internal manifolds
are turned on.   The Dirac quantization condition tells us that
these fluxes obey integer quantization rules, so we anticipate a
large discrete lattice of solutions, for an internal manifold of
complicated topology.  A particularly simple set of solutions was
found by \cite{GKP} using the Lagrangian of Type IIB SUGRA. The
internal manifold is conformal to a Calabi-Yau manifold, there are
imaginary self dual fluxes of a combination of the Ramond-Ramond and
Neveu-Schwarz 3-form fields.  The flux superpotential fixes all the
complex structure moduli and string coupling.   The Kahler moduli
remain moduli of these solutions.  For some choices of fluxes, the
fixed value of the string constant is numerically small, so one
claims that one can still trust notions from weak coupling string
theory.

One feature of this system that is quite general is the necessity
for an orientifold in addition to classical super-gravity fields.
All weak coupling string theory approaches to compactification will
have to deal with the dilaton field.   Apart from Calabi-Yau
compactifications with no flux, (for which the Einstein Lagrangian
vanishes on shell), there will always be sources for the dilaton
field in the compact dimensions. The classical SUGRA contributions
to the dilaton source are all positive, so we get an equation
$$ \nabla^2 \phi = P.$$  Integrating this equation over a compact
manifold, we get a contradiction\cite{teninto4wontgo}.  The
orientifold provides a negative source term, which allows for
consistent solutions. Orientifolds are singular and do not belong in
effective field theory, but they are certainly innocuous in weakly
coupled string theory in flat space.  As long as one has the weak
coupling string theory formalism at one's disposal, one can imagine
that this remains true in curved space. Orientifolds can be defined
in a finite manner if one has a world sheet sigma model.  We'll
discuss this further below.

At tree level, the solutions preserve SUSY in Minkowski space if the
value of the flux superpotential at the minimum vanishes.  Other
choices of fluxes, for which $W_0 \neq 0$ break supersymmetry.
These solutions still have vanishing cosmological constant because
the Kahler potential of Type IIB SUGRA for the Kahler moduli, has
the so called no-scale form. However, quantum corrections to the
Kahler potential or superpotential will change this. While the
latter is exponentially small in the compactified Kahler moduli one
can argue that for small $W_0$ it can still be the dominant effect
at large Kahler moduli.  One finds (AdS)-supersymmetric solutions
with negative c.c. by tuning fluxes so that $W_0$ is small.

There are, in my opinion, two related things to worry about in these
solutions.  The first is the question of what it means for the
string coupling to be small, and the second is what to do about the
orientifold.  The normal meaning of small string coupling is that
there is a world sheet expression for observables, and an expansion
in powers of the string coupling by summing over world sheet
topology.   If there were such an expansion, we would have no
problem defining orientifolds as finite world sheet field theories.
But there cannot be such an expansion in this context, because the
string coupling is fixed by competition between different terms in
the $g_S$ expansion of the superpotential.  So we must view these
compactifications as constructs in effective field theory, but the
orientifold is problematic in a long wavelength expansion.  We know
that orientifolds are perfectly finite in flat space perturbative
string theory, and many orientifolds are related by dualities to
smooth solutions of M-theory. So it is not so much the existence of
the orientifold that is at issue, but rather whether its singularity
could hide dependence on the fluxes which are the control parameters
for these solutions.

Many of these ambiguities are removed, at the expense of a
considerable loss in computational power, by looking at F-theory
compactifications.  F-theory is a rubric for a class of solutions of
Type IIB string theory, in which the complex string coupling $\tau =
a + i\frac{4\pi}{g_S}$ ($a$ is the RR axion field) varies over a
complex 3-fold base space of large volume.  The ensemble defines an
elliptically fibered $CY_4$ space, with $\tau$ describing the
complex structure of the elliptic fiber. The orientifold solutions
described above are special limits of F-theory compactifications,
which were introduced in order to use weak coupling methods. In a
general F-theory compactification the string coupling varies over
the 3-fold base and is never weak everywhere.  In the orientifold
limit the region where the coupling is strong shrinks to the locus
of the singular orientifolds.  More generally, the only expansion
parameter in F-theory is the volume of the 3-fold base in 10
dimensional Planck units. F-theory models with fluxes also exist and
have been studied extensively in recent years\cite{Fflux} .  While
the flux induced superpotential for the complex structure moduli of
the base has not been computed explicitly, there seems little doubt
that for sufficiently generic fluxes all the moduli will be fixed,
leaving only the Kahler moduli. For simplicity we can assume that
there is just one Kahler modulus. There is at least one since the
overall volume of the compact space will not be determined by the
SUGRA action.

Thus it is extremely plausible that on a 3-fold base with large
$h_{2,1}$ there will be a large number of smooth solutions of Type
IIB SUGRA, with all moduli but the overall volume fixed.  Below the
Kaluza- Klein scale there will be an effective four dimensional
theory with $N=1$ SUGRA and a single chiral multiplet with a
no-scale superpotential.  The superpotential $W_0$ will be a flux
dependent constant, and since there are many fluxes, it is plausible
that it can be tuned to be much smaller than the KK energy scale, as
one finds explicitly for the superpotential computed in the
orientifold limit. The use of classical SUGRA is of course
predicated on the assumption that the KK radius is much larger than
the ten dimensional Planck length.  These solutions preserve $N=1$
SUSY only if the superpotential vanishes. (One way to guarantee this
is to search for solutions that preserve a discrete R symmetry. The
volume modulus will have R charge 0.) However, as a consequence of
the no-scale Kahler potential, all of them will have four flat
Minkowski
 dimensions.

This is not consistent if $W_0 \neq 0$.  If it were, there would be
a low energy effective action for the modulus in four dimensional
$N=1$ SUGRA, but corrections to the Kahler potential would change
the cosmological constant, and there could not be a Minkowski
solution.   However, it does make sense to postulate the existence
of a supersymmetric AdS solution. The condition for supersymmetry is
$$\partial_\rho W - \frac{1}{m_P^2} \partial_\rho K W, $$ where we
have parametrized the Kahler modulus as
$$ (R m_{10})^{-4} = \frac{{\rm Im}\ \rho}{m_P}.$$
The real part of $\rho$ is an angle variable, so all corrections to
the superpotential must be integer powers of $e^{2\pi
i\rho}$\footnote{If, in F-theory, the cycle associated with this
Kahler modulus is wrapped by multiple seven branes, so that the
$CY_4$ is singular on that cycle, then the shift symmetry of the
angle variable induces a chiral transformation on the matter fields
that couple to the 7-brane gauge group.  This chiral symmetry is
spontaneously broken by strong coupling gauge theory dynamics, which
introduces a new discrete finite variable into the effective field
theory, parametrizing the different field theory vacuum states. The
result is fractional powers of $e^{2\pi i \frac{\rho}{m_P}}$ in the
effective superpotential.  Unless the 7-brane group is very large,
this does not substantially change our argument.}.   Following KKLT
one can then argue that if $W_0$ is small, the system is
self-consistently stabilized at a large value of the imaginary part
of $\rho$, where the corrections to the classical Kahler potential
are small.

Our own analysis of this situation differs from that of KKLT in two
ways.  Rather than considering it a controlled calculation in string
theory, we view it as a plausible self consistency check for the
existence of a supersymmetric AdS model of quantum gravity, whose
low energy Lagrangian and background configuration are those
suggested by KKLT.   The second difference is that we reject the
idea that the weak coupling orientifold calculation is more
controlled than the general F-theory set-up.  The former has an
orientifold singularity, whose effect can only be estimated if we
have a systematic world sheet expansion.  However, the model fixes
the string coupling at a value that is not parametrically small, so
no world sheet calculation is likely to exist.   The only world
sheet calculation one can attempt is an expansion around one of the
Minkowski solutions of the classical string equations with the
orientifold source.  We know that if $W_0 \neq 0$, the string loop
expansion leads to divergences in the integral over toroidal moduli
space.  One can attempt to cancel these divergences with the
Fischler-Susskind mechanism\cite{fs}, but this leads to a time
dependent background, which is singular in either the past or the
future or both.  It does not correspond to the stable supersymmetric
model whose existence we are asserting.

The only real calculational advantage of the orientifold limit of
F-theory is the exact formula for the flux induced superpotential.
Rather than pursuing the idea that weak coupling string perturbation
theory can be used to calculate some useful property of the
hypothetical supersymmetric AdS model, it would seem more profitable
to try to find an analogous formula for the superpotential in
general F-theory models, or at least to argue that a general model
with generic fluxes will indeed stabilize all the complex structure
moduli.

The bottom line of this discussion is that F-theory
compactifications with generic fluxes seem to stabilize all complex
structure moduli at the level of classical SUGRA.  We use the phrase
seem to because detailed calculations rely on the GVW
superpotential, calculated at weak string coupling. Even in the
orientifold limit of F-theory, there is no systematic string loop
expansion of these models, when $W_0 \neq 0$. Classical solutions in
which $W_0 = 0$ as a consequence of an anomaly free discrete R
symmetry provide us with moduli spaces of asymptotically flat models
of quantum gravity in four dimensions. The Kahler moduli are exact
moduli of these models.   When $W_0 \neq 0$ we have, at the
classical SUGRA level SUSY violating asymptotically flat solutions.
The classical SUGRA equations are formally exact in the limit of
infinite Kahler moduli, but corrections to this approximation ruin
asymptotic flatness.  {\it If we assume the existence of large
radius, supersymmetric AdS compactifications}, then effective field
theory provides a self consistent solution.  Such compactifications
should have a low energy effective field theory, and symmetries
constrain the superpotential and Kahler potential of that field
theory. If $W_0 \ll 1$ (in Planck units) then we find a self
consistent solution in which we only need to include the leading
correction to the superpotential. For F-theory solutions whose
$CY_4$ has large Betti numbers, of order $100$, there are many
solutions of this type and it seems plausible that one can find many
examples with small $W_0$. The effective expansion parameter is
$|{\rm ln} \frac{W_0}{m_{10}^3}|^{-1} $, and the scale of
Kaluza-Klein excitations is parametrically larger than the inverse
AdS curvature radius.  It is believed that the Betti numbers of
$CY_4$-folds are bounded, so the expansion parameter can never be
really small.

The KKLT paper can thus be viewed as providing evidence for a large
class of large radius supersymmetric AdS compactifications.  This
conjecture is subject to a rigorous test.  One must find $3$
dimensional superconformal field theories whose properties mirror
those of the conjectured geometries via the AdS/CFT correspondence.

DeWolfe {\it et. al.}\cite{dewolfe} have suggested another set of
supersymmetric AdS solutions with a tunably small parameter.  These
are based on solutions of (massive) Type IIA string theory. They
again purport to have small string coupling and a parametrically
suppressed ratio between the compactification radius and the AdS
radius. In these compactifications, the control parameter is a large
flux, $N$. However, in \cite{vandenbroek} we provided evidence that
the ever-present orientifolds in such weak coupling constructions
hide a region of the compact manifold where the string coupling is
large and the compactification radius scales like the AdS radius.
The picture in \cite{vandenbroek} provides an explanation for the
scaling of the entropy with $N$, which is not available in the weak
coupling picture. Again, the real test of all of these conjectures
is to find superconformal field theories with the properties implied
by these geometries.  This is particularly interesting in the Type
IIA case, because \cite{dineetal} have exhibited non-supersymmetric
versions of these compactifications, which look equally plausible.
This implies the existence of a large class of non-supersymmetric
fixed points with large radius AdS duals.  As I've emphasized above,
neither orbifolding nor holographic RG flow, both of which seem like
plausible mechanisms for finding examples of such large radius CFTs,
actually succeed.

It seems to me that this is a place where progress can be made in
assessing the reliability of the effective field theory approach to
the String Landscape.  There is an apparent conflict between the
vast landscape of SUSY violating large radius AdS duals promised by
the construction of approximate effective potentials, and our
inability to construct even one example of the same from a reliable
starting point.  Perhaps the most controlled setting for studying
this problem is that of $AdS_3$ models.  The effective potential
approach to these is quite similar to that for $AdS_4$, but in $2$
dimensions we have a much richer arsenal of tools for studying CFTs
without recourse to perturbation theory.  This area is relatively
unexplored and might repay the attention of young researchers.

I've deliberately avoided discussing the procedure of ``uplifting
the AdS solutions to meta-stable dS solutions by adding
anti-D3-branes".  This purports to be a small perturbation of the
existing solutions, but it is manifestly not.  No one knows how to
describe the observables of meta-stable dS states, but it is clear
that they have nothing to do with conformal field theories living on
the boundary of a $3+1$ dimensional AdS space.   The procedure of
adding anti-branes is perfectly sensible when we are talking about a
brane configuration of non-compact codimension 3 or more, embedded
in a string model in asymptotically flat space-time.  It may also be
valid in co-dimension 2.  For co-dimension zero the back reaction of
branes on the geometry is simply not a small perturbation.  If we
recall that even a small change in the c.c. changes the high energy
spectrum of the theory, we see immediately that one cannot play
perturbative or low energy effective field theory games in this
situation. We will discuss a possible theory of meta-stable dS
spaces below.

\subsection{Tunneling in gravitational theories}

The key paper on gravitational tunneling is that of Coleman and De
Lucia\cite{cdl}.  I urge every serious student of this subject to
study that paper carefully and completely.  The study of tunneling
in general quantum systems is the study of {\it instantons}:
Euclidean solutions of the classical equations of motion {\it with
appropriate boundary conditions.}  In QFT in Minkowski space, the
boundary condition is that the scalar fields must rapidly approach
their values at some meta-stable minimum of the scalar potential, as
the radius goes to infinity.  The classical solution is $O(d)$
symmetric in $d$ Euclidean space-time dimensions, and defines a
finite ``critical bubble". The bubble wall is generically fuzzy, and
is defined by saying that the field is closer to the meta-stable
minimum than some small parameter $\epsilon$. The derivative of the
scalars vanishes at the center of the bubble, and this allows us to
analytically continue the bubble geometry to the interior of a
forward light cone in Minkowski space. The Euclidean solution
provides initial conditions for the propagation of the scalar field
inside this light cone.  It is easy to see that as one proceeds
forward on homogeneous slices of constant negative curvature the
scalars smoothly approach their values at the absolute minimum of
the potential. One says that {\it the false vacuum has decayed into
the true vacuum}.  We will continue to use the terms true and false
minimum in the gravitational case even though we have emphasized
that the concept of vacuum state does not make any sense in quantum
gravity.  We will also see that not all instantons describe decay.

In finite temperature field theory, this prescription is modified.
The Euclidean time dimension is compactified on a circle, and one
searches for periodic Euclidean solutions. The solutions no longer
achieve the false minimum, and they describe the decay of a
meta-stable thermal ensemble, through a combination of quantum
tunneling and thermal hopping over the barrier.

Coleman and De Lucia generalized this prescription to include the
dynamics of the gravitational field. Their presentation is oriented
towards situations where the gravitational effects are ``a small
perturbation" of the flat space theory, but they discovered that in
many cases this claim is untenable, and the gravitational effects
are large. We will not make such a restriction, but it's important
to emphasize that CDL discovered examples of every phenomenon we
will discuss, within the confines of their restricted
approximations. One of the most important features of the CDL
analysis is the way in which the nature of gravitational tunneling
depends on the cosmological constants at the true and false minima.
We will present this as evidence that the nature of the actual
quantum theory is in fact quite different in the case of zero,
positive and negative c.c. .

People often ask me why I place so much confidence in the CDL
calculations, since I am always warning that too much reliance on
the field theory approximation is dangerous.  Indeed, in the
proposals I will present below the metric of space-time is not a
fluctuating quantum variable, but is instead determined by a rigid
set of kinematic constraints on the quantum theory.  I believe a
reasonable analogy is presented by the Wilson loop variables of
large N gauge theory.  In the planar limit, the Wilson loop
expectation value satisfies a classical field equation in loop
space\cite{migmak} and the $1/N$ expansion can be viewed as a sort
of Feynman diagram (string loop) expansion around this classical
equation.  However, for finite $N$ the Wilson loop operators are not
independent canonical variables, and the Hilbert space of the
perturbation expansion is too big.  The true quantum variables are
the gauge potentials in some physical gauge. Nonetheless, Euclidean
solutions of the equations for Wilson loops can be used to find
tunneling corrections to the $1/N$ expansion.  However, the real
reason for paying attention to the CDL results is that they can all
be related to more fundamental concepts in the theory of QG;
concepts like the holographic principle and the AdS/CFT
correspondence.  We will now proceed to classify gravitational
tunneling events according to initial and final values of the c.c. .

\subsection{No tunneling to or from AdS space}

One of the most annoying aspects of this subject is the tendency of
many speakers to talk about tunneling to AdS space.  Perhaps the
most important point in the CDL paper is the demonstration that this
NEVER occurs, except in the thin wall approximation.  To understand
the result we write the CDL equations for the gravitational field
coupled to a scalar via the Lagrangian $${\cal L} = \sqrt{-g} [R -
\frac{1}{2} (\nabla \phi )^2 - V(\phi )]$$. We work in four
dimensions, with a single field, for simplicity, but our conclusions
are general. Given a scalar potential

$$V(\phi ) = \mu^4 v(\phi / M),$$ the natural space-time scale for motion
is $L = \frac{M}{\mu^2 }$.  If we make a Weyl transformation to
dimensionless field variables (we use conventions where coordinates
are dimensionless and the metric tensor has dimensions of squared
length), and write an $O(4)$ symmetric ansatz:

$$ ds^2 = L^2 (dz^2 + \rho^2 (z) d\Omega^2 ),$$ $$\frac{\phi}{M} =  x (z),$$
where $z$ is a dimensionless radial coordinate and $\rho $ is the
dimensionless metric coefficient, then we get Euclidean field
equations

$$ (\rho^{\prime})^2 = 1 + \epsilon^2 \rho^2 [\frac{1}{2}
(x^{\prime})^2 - v(x) ] .$$  $$x^{\prime\prime} + 3
\frac{\rho^{\prime}}{\rho} x^{\prime} = \frac{dv}{dx} .$$ $\epsilon
=  \frac{M}{\sqrt{3} m_P}, $ where $m_P$ is the reduced Planck mass
$2 \times 10^{18}$ GeV.  Note that although $\mu$ does not appear
explicitly in these equations, it must be less than $m_P$ for the
semi-classical approximation to be valid (how much less is a matter
of conjecture).  Note also that the quantity in square brackets in
the first equation is what would have been the ``conserved energy"
of the second equation in the absence of the friction term.

By convention, the center of the bubble is at $z=0$ and in the
vicinity of this point $\rho = z$.  The solution is non-singular
only if $x^{\prime} (0) = 0$.  The boundary condition at the upper
end of the $z$ interval depends on the c.c. in the false minimum.
Our present considerations are independent of that boundary
condition. To analytically continue the solution to Lorentzian
signature we take $z = it$ and use the Euclidean solution at $z=0$
as an initial condition for the Lorentzian evolution.  The initial
conditions are $\dot{x} (0) = 0$, $\rho (0) = 0$ and $x(0)$ a fixed
value determined by the boundary conditions at the other end.  It
must be in the basin of attraction of the true minimum.

The Lorentzian equations are
$$\dot{\rho}^2 = 1 + \epsilon^2 \rho^2 [\frac{1}{2}\dot{\phi}^2 +
v].$$

$$\ddot{x} + 3\frac{\dot{\rho}}{\rho}\dot{x} + \frac{dv}{dx} = 0 .$$

These equations have an AdS solution in which $x$ is equal to the
true minimum of $v$ for all time, and $\rho = \sin
(\sqrt{\Lambda}t)$. However, the solution determined by the
instanton does not approach this solution, which is unstable to
infinitesimal perturbations which are homogeneous and
isotropic\footnote{AdS spaces are sometimes stable to small
perturbations which fall rapidly at infinity.  These are the
normalizable fluctuations of the AdS/CFT correspondence.   The
homogeneous isotropic solutions relevant for instanton physics are
not normalizable. Generally they have no extension outside the FRW
coordinate patch, as a consequence of the singularity we are
discussing.}.  Indeed, since the Euclidean solution completely fixes
the initial conditions for Lorentzian evolution, $\dot{x}$ will not
go to zero as $\rho \rightarrow 0$.  The kinetic energy of $x$ goes
to infinity, because the universe is contracting and we have Hubble
anti-friction. $x$ will not stay near the true minimum, but will
explore its whole potential surface.  This singularity will be
reached in a time of order $\frac{M}{\mu^2 \epsilon}\sim \frac{m_P
}{\mu^2}$.  In a typical particle physics model $\mu$ is unlikely to
be smaller than a few hundred MeV, so this time is shorter than
$10^{-5}$ sec. For future reference we note that, according to the
holographic principle, this implies that an observer trapped in this
region can access an entropy that is at most $\sim
(\frac{M_P}{\mu})^4 < 10^{80}$, only $\sim 10^{60}$ of which can be
in the form of matter and radiation.  The actual matter/radiation
entropy of our universe is $\sim 10^{80}$. The reader who is
confused by these numbers, will be able to go back and check them
after we discuss the holographic principle.

The converse of this result is also true: a quantum AdS space cannot
decay by tunneling.  This follows from the AdS/CFT correspondence.
The exact mathematical formulation of CFT requires one to have only
unitary highest weight representations of the conformal group in the
Hilbert space.  It follows that the global Hamiltonian $K^0 + P^0$
is bounded from below.  But the Lorentzian continuation of an
instanton is always a zero energy solution in which the positive and
growing kinetic energy of the expanding bubble is balanced by the
increasingly negative potential energy of its interior.  It always
corresponds to a system which is unbounded from below.

This general argument is exemplified in a beautiful paper by Hertog
and Horowitz\cite{herthor}.  These authors found an instanton
solution, which seemed to indicate a non-perturbative decay of the
supersymmetric(!) $AdS_4 \times S^7$ solution of 11 dimensional
SUGRA.  Upon closer examination, they found that although the
perturbation fell off at infinity, it was not a normalizable
solution, corresponding to a state in the CFT.  Rather, it
corresponded to a perturbation of the CFT by a marginal operator
that was unbounded from below.  The instantons for the putative
decays of AdS vacua have also been studied
by\cite{eliezer}\cite{harlow}, who provide further evidence that the
landscape interpretation of these events is faulty.

The correct way to interpret these facts is to say that if we look
at a classical bulk Lagrangian, which has an AdS solution, as well
as an instanton which behaves like a normalizable perturbation of
this solution at Euclidean infinity, then we will have proven, in
the classical approximation, that the AdS Hamiltonian of this system
is unbounded from below, and cannot have a CFT dual.  It is likely
that such a solution is not part of any sensible quantum theory of
gravity.  Indeed, there is an interesting sidelight on this
situation, which already indicates that something serious is wrong
with the interpretation of this instanton as a decay of the original
AdS space-time.

In ordinary quantum field theory, excitations around the false
vacuum are meta-stable only up to some finite energy.  If we make
the energy density larger than the barrier height the system is
simply unstable.  Similarly, the thermal ensemble is meta-stable
only up to some finite temperature.  In quantum gravity in large
radius AdS space, we can explore the thermal ensemble by looking at
AdS-Schwarzschild solutions of the field equations, which are
normalizable and have positive energy of arbitrarily large size.
These solutions do not have classical instabilities, indicating that
the vacuum decay paradigm of non-gravitational QFT is breaking down.

We can gain more insight into this when we realize that the
expanding bubble of the Lorentzian instanton {\it does not penetrate
the interior of a black hole}.  The bubble expands only at the speed
of light, while the interior geometry expands away from the bubble
superluminally.  A solution whose initial conditions consist of a
space-like separated pair of a black hole and a nucleated critical
bubble, has two causally separated future asymptotic regions, both
of them space-like singularities. Multiple black holes in the
initial state will lead to multiple causally disconnected future
regions. Furthermore, single the bubble nucleation probability is
{\it exponentially small} as $\frac{m_P}{\mu}$ goes to infinity, it
is easy to see that the black holes can have exponentially larger
entropy than the entropy accessible within the bubble.  These
semi-classical considerations suggest very strongly that there is no
sensible quantum mechanical interpretation of AdS solutions that
have genuine instanton instabilities.  Certainly the interpretation
of the instanton as a decay of the original AdS ``state" into the
system in the interior of the CDL bubble, is completely untenable.
This analysis goes through in precisely the same way for CDL
``unstable" asymptotically flat space-times, although the existence
of Hawking instabilities of black holes in that case, poses further
complications.

Our conclusion is that AdS solutions of bulk gravitational field
equations never arise as the result of CDL decays, and do not decay
in a way that resembles the vacuum decay of a non-gravitational QFT.
Some of the solutions are stable, and may well belong to a real
theory of QG, which would be defined by a CFT dual.  The unstable
ones surely belong to a very peculiar quantum theory, if they have
any meaning at all.  There is thus strong evidence from CDL
tunneling, complementing that from the AdS/CFT correspondence, that
AdS solutions of gravitational field equations form little isolated
models of QG, which have nothing to do with a larger landscape.

\subsection{Gravitational tunneling to and from zero c.c. states}

In asymptotically flat space-time, the asymptotic symmetry algebra
is the Poincare group.  If we do not insist on
supersymmetry\footnote{It is one of my contentions that we MUST
insist on SUSY {\it i.e.} that every asymptotically flat model of QG
is in fact Super Poincare invariant.  However, we are exploring more
general possibilities in this section, and our explorations lead to
important insights for the program based on my conjecture.} there is
no general argument that the Hamiltonian is bounded from below.
However, there is a classical theorem\cite{posenthm} which shows
that asymptotically flat solutions of certain Lagrangians do have
classically positive energy.

The paper \cite{abj} clarified how the space of theories consisting
of scalar fields coupled to gravity is divided up by the positive
energy theorem.  Consider a potential with classically stable
Minkowski and AdS solutions and ask whether there is a static domain
wall connecting the two solutions. For the AdS/AdS case, we saw that
such domain walls with boundary conditions that correspond to
normalizable solutions on both sides of the wall, are the
holographic representation of RG flows between two fixed points. No
such interpretation is possible here, because the analog of the UV
fixed point is the Poincare invariant model, which is not a quantum
field theory.  The equations determining the domain wall are

$$ \rho^{\prime\ 2} = \epsilon^2 (\frac{1}{2} x^{\prime\ 2} + v(x))  $$
$$ x^{\prime\prime} + \frac{{3 \rho^{\prime}}}{\rho} x^{\prime} +\frac{dv}{dx} = 0  ,$$
with boundary conditions
$$  x(\pm\infty) = x_{\pm} .  $$ $x_{\pm}$ are the false and true
maxima of $v(x)$.

As is familiar from linear eigenvalue problems, this system does not
have solutions for a generic potential.  In the limit in which we
model the domain wall as an infinitely thin brane with a given
tension, there will only be one value of the tension for which the
static solution exists\cite{cvetic}.  For tensions below this value
there is instead a solution which looks like the asymptotic limit of
an expanding bubble wall, corresponding to CDL decay of the
Minkowski background (but missing the instability of the previous
subsection, for which one must go beyond the thin wall
approximation). For tensions above this there is no interpretation
of the solution as the limit of an object in the Minkowski
background.

More generally, as in any eigenvalue problem, we can find a solution
obeying both boundary conditions by tuning a single parameter in the
potential.  {\it Thus, the space of all potentials with a Minkowski
solution of the field equations contains a co-dimension 1
submanifold, on which a static domain wall connecting Minkowski
space to one particular AdS minimum exists, while all for all other
AdS minima there are neither domain walls nor expanding bubble
solutions. } For a given Minkowski minimum there will generally be
only one domain wall, though in supersymmetric situations there may
be more \footnote{In the AdS to AdS case, static domain walls
correspond to relevant perturbations of the CFT for smaller absolute
value of the c.c. , which point along RG flows to other fixed
points. Such flows are non-generic unless both fixed points are
supersymmetric.}.  This submanifold in the space of potentials is
called {\it the Great Divide}.  By perturbation theory one can show
that above the Great Divide the Lagrangian has a positive energy
theorem, while below it there are expanding bubble solutions and the
ADM energy is unbounded from below.

In \cite{abj} we showed that by varying the parameter $\epsilon \sim
M/m_P$ in potentials of the form $\mu^4 v(\phi / M) $ we crossed the
Great Divide.  For $\epsilon \ll 1$, the non-gravitational analysis,
which indicates an instability is essentially correct.  However, it
should be noted that even in this regime, CDL showed that one is
above the Great Divide if $|v(x_T)|$, the magnitude of the c.c. at
the true minimum, is $\ll 1$.  The Great Divide itself is located at
$ \epsilon = o(1)$ for generic functions $v(x)$.  For those models
below the Great Divide, the same issues arise as for unstable AdS
spaces. Starting from a generic excited state of the Minkowksi
solution, we end up with a future that contains multiple causally
disconnected space-like singularities, most of whose entropy is
contained in black holes.  Here however we have to deal with the
perturbative Hawking instability of black holes, which returns the
degrees of freedom of the black hole to a region causally connected
to the expanding bubble.  Here we can encounter a paradox: The
matter entropy outside the bubble is bigger than that measurable by
any observer inside the bubble.   One suspects that we are being too
naive and neglecting back reaction of all of this matter on the
bubble.  A possible scenario is that collisions of the bubble wall
with a sufficiently large matter density, converts the bubble into a
black hole. Indeed, the bare expanding bubble solution has exactly
zero energy in empty space.  If it collects a finite surface energy
density as it passes through a region filled with a uniform density
of matter, then it will end up with a mass of order the square of
its radius.  For large enough radius the Schwarzschild radius of
this distribution will be larger than the bubble radius.  Thus, a
resolution of the apparent paradox of a bubble sweeping up more
entropy than any observer inside it can measure, may simply be that
in attempting to swallow all of this entropy, the bubble forms a
black hole around itself.

The bizarre conclusion of this story would be that, perhaps, below
the Great Divide, empty flat space is unstable, but flat space with
enough entropy in it nucleates a black hole around the expanding
bubble.  Of course, another possibility is that there are no actual
theories of quantum gravity which contain such meta-stable flat
space-time configurations.  When we discuss the holographic
space-time formalism, we will show that it suggests that all quantum
theories of asymptotically flat space-time are exactly
supersymmetric.  If this is the case then they are automatically
Above the Great Divide.  This does not yet settle the question of
the fate of the asymptotically dS universe, which we appear to
inhabit.

\subsection{CDL transitions from dS space}

If we take a potential below the Great Divide and add a small
positive constant to it, we do not make a significant change in the
CDL transition rate.  The entire story of the previous section
replays with little change.  Above the Great Divide the story is
different.  With mild assumptions, there is {\it always} a CDL
instanton when a potential has a positive and negative minimum
separated by a barrier\footnote{The exceptions come for potentials
in which the maximum is very flat.  Analogies with ordinary quantum
mechanics lead us to expect a transition from the false minimum to
the top of the barrier, which is more or less semi-classical,
followed by large quantum fluctuations on the flat top.  However,
since the system includes gravity, we don't really know how to
explore the regime of large quantum fluctuations.  It is possible
that potentials this flat are simply forbidden in real theories of
QG\cite{bdfg}\cite{amnv}\cite{bjs}.}. However, the transition rate
goes to zero like $$e^{ - \pi (RM_P)^2},$$ as $R$, the dS radius
goes to infinity. This suppression by something of order the inverse
of the exponential of the dS entropy, is what we would expect for a
transition at infinite temperature for a system with a large finite
number of states, into a very low entropy state.  This is consistent
with our previous remark that, according to the holographic
principle, the maximal entropy observable in the negative c.c. Big
Crunch is a microscopic number.  We will see below that the
interpretation of dS space as a system with a finite number of
states, at infinite temperature, is consistent with all
semi-classical evidence about dS space, including its finite
Gibbons-Hawking temperature!

Transitions from one dS space to another are also consistent with
this picture, and add an extra bit of evidence.  Indeed, although we
have not emphasized it above, the instantons for transitions out of
dS space are compact manifolds, with positive scalar curvature, just
like Euclidean dS itself.  And like Euclidean dS space they have
negative action. The probability interpretation of the instanton
calculation comes by subtracting the dS action from the instanton
action, which always gives a positive number. In the case of dS to
dS transitions, we get two different probabilities, depending on
which dS action we subtract.  These are interpreted as the
probabilities for the forward and reverse transitions

$$P_{1\rightarrow 2} = e^{- (S_I - S_1)} , $$ $$P_{2\rightarrow 1} =
e^{- (S_I - S_2)} .$$ The ratio of transition rates is thus
$$e^{ - (S_1 - S_2)}.$$ It is a quite remarkable fact (analogous to
a result about black holes first discovered by Gibbons and Hawking),
that the dS action is exactly the negative of the dS entropy. This
means that these transition rates satisfy the principle of detailed
balance appropriate for a system with a finite number of states at
infinite temperature. Unlike the case of dS transitions to a
negative c.c. Crunch, this semi-classical calculation is under
control in both directions.  It seems perverse to attach any other
meaning to it than what it seems to say: dS space is a system with a
finite number of states.  Its Hamiltonian is generic and the time
evolution of a randomly chosen initial state will sweep out the
entire Hilbert space. The dS space with larger c.c. is a low entropy
configuration of this system and will be accessed only rarely, in
direct proportion to the fraction of the total number of states
corresponding to this configuration.

Note that this interpretation meshes perfectly with the one we have
proposed for dS to Crunch transitions above the Great Divide. Note
further that it does not agree with ANY interpretation of the same
transition according to the theory of Eternal Inflation.

As we take the smaller c.c. to zero, the transition rate to the
higher c.c. state goes to zero. This makes sense in our
interpretation, because the probability of finding a finite entropy
subspace of states starting from a random search through an infinite
dimensional Hilbert space, is zero.  Note however that the limit of
zero c.c. is a very subtle one.  In the section on stable dS space,
we will see that a lot of states must be discarded from the dS
Hilbert space in order to describe the Hilbert space of the limiting
Poincare invariant theory.  The entropy of the latter scales as
$(RM_P)^{3/2}$ as the dS radius goes to infinity, while the total
entropy of the dS Hilbert space is $\pi (RM_P)^2$.  We will see in
the next subsection that the required limit for a tunneling solution
whose target is a zero c.c. space-time is quite different. The
interpretation of such solutions is intertwined with attempts to
construct a theory of the String Landscape, and we turn to that
problematic subject next.

\subsection{Implications for the landscape}

The implications of these results in semi-classical gravity for the
idea of a string landscape are profound.  Asymptotically flat and
AdS models of quantum gravity {\it are not part of the landscape}
and do not communicate with hypothetical landscape states by
tunneling.  Tunneling only makes sense for meta-stable dS points on
an effective potential. These can tunnel to other dS points, to
negative c.c. Big Crunches, and to zero c.c. states.  None of the
physics of these states is encoded in anything like the boundary
correlators that string theory has taught us how to compute.  {\it
If the landscape exists, the very definition of its observables must
be completely different from that of ordinary string theory.}

We have seen that tunneling to negative c.c. crunches falls into two
categories.  Above the Great Divide, we've provided a plausible
quantum interpretation of the CDL tunneling probabilities, in terms
of a quantum theory of stable dS space with a finite dimensional
Hilbert space.   Below the Great Divide, we've argued that these
transitions are fraught with interpretational ambiguities.  The true
endpoint of CDL decay is not a quiescent true vacuum, nor even a
single big crunch.  The final state depends on which initial excited
state of the dS or flat ``false vacuum" one begins with.  It
typically has multiple crunching regions, with different pre-crunch
internal geometries, which are causally disconnected from each
other.

Nonetheless, many advocates of the landscape insist that any
sensible meta-stable model of dS space must be below the Great
Divide. The argument is somewhat philosophical, but depends
crucially on the fundamental claim that the landscape solves the
c.c. problem by invoking the anthropic principle.  In order to be
certain that this is true, one counts meta-stable landscape points,
according to some criterion, and claims that the number is of order
$10^{500}$ or greater.  It is important that this number is {\it
much larger} than the ratio between a Planck scale c.c. and the c.c.
we observe.  One then argues that if generic minima of the potential
have a c.c. that is a sum of a such a large number of positive and
negative Planck scale contributions\footnote{In the
Bousso-Polchinski\cite{bp} version of this argument there is one
large negative contribution and a large number of smaller positive
ones.  In KKLT, one argues for a set of AdS solutions, with enough
free parameters to make the negative c.c. small, and then adds a
small positive contribution.}, there will inevitably be some with
c.c. of the value we observe. If anthropic arguments can show that a
value bigger than this is incompatible with the existence of
intelligent life forms, one has ``explained" the small value of the
c.c.   Note that in order for this counting to work in a way that
does not require close scrutiny of each and every minimum of the
potential, there must be MANY solutions with a value of the c.c.
close to ours. There is no reason for other properties of the low
energy world to be similar to those we see. So questions like what
the low energy gauge group and representation content are, as well
as the value of most low energy parameters, must also be answered
anthropically.

I will not spend time here rehashing the futility or experimental
implausibility of this claim, but rather emphasize the general
picture of string theory that it implies: string theory has MANY
solutions with small c.c.  If we are not to regard our own world as
simply an accidental consequence of the theory, then we must come up
with some argument that makes our conditions more hospitable for
observers than other possible meta-stable states.  The advocates of
these ideas are led to contemplate the question of whether we are
{\it typical} members of the class of observers that the landscape
predicts.   The answer is that this can only be true if our current
dS condition does not last too long.

We know that the universe we observe began in a state of much lower
entropy than it has today.  This is why we see the second law of
thermodynamics in operation. The visible entropy of the universe is
dominated by cosmic microwave background photons, and the total
entropy by the supermassive black holes in the centers of galaxies.
According to modern cosmology, this entropy was created in the
post-inflationary history of the universe, through the decay of the
inflaton field into radiation, and the gravitational collapse caused
by the action of the fluctuations of this field on non-relativistic
matter.  In the landscape picture, the beginning of this cosmic
history is a tunneling event from a higher c.c. meta-stable point,
to our own basin of attraction.  It is a very low entropy
fluctuation.

If the current c.c. dominated era of the universe lasts too long,
there is a much more efficient way to make observers than to have a
fluctuation that recreates the entire history of the universe.  Such
a fluctuation, by the CDL calculation, has a probability of order
$e^{ - 10^{123}}= e^{ - A/4}$, where $A = 4 \pi (RM_P)^2 $.  On the
other hand, in the asymptotic future dS space, the probability to
have a random fluctuation that creates a localized mass equal to
that of a ``single intelligent observer" is $e^{ - 2 \pi R m_O}$ and
the probability that that mass is in the state corresponding to a
live intelligent observer is at least $e^{ - \frac{A_O}{4}}$, where
$A_O$ is the horizon area of a black hole which could enclose the
observer. These ridiculously tiny probabilities, are much larger
than the probability of the fluctuation that started the universe
off. So, either the landscape explanation of the origin of our
universe is wrong, or we are far from typical observers, {\it or the
dS state must decay long before all these typical observers can be
formed.} This is only possible if our meta-stable dS state is below
The Great Divide, which is the choice made by many landscape
theorists. As we will see, this claim creates some tension with the
only extant proposal for making a true theory of the string
landscape.

The attempt to create a true theory of the landscape, analogous to
our models of asymptotically flat or AdS spaces has been centered
entirely in the Stanford-Berkeley group.  The proposal is that the
observables of the theory somehow reside in the causal diamond of a
post-tunneling event into a zero c.c. region of the potential, which
locally approaches one of the maximally supersymmetric flat space
solutions of string theory.  The original idea was to construct a
sort of scattering theory in the Lorentzian space-time defined by
the CDL instanton.  It's indeed true that if we consider quantum
field theory in such a space-time, one can define scattering states
on the past and future boundaries.   However, as I suggested in
2004, and was later proven rigorously by Bousso and Freivogel,
generic scattering boundary conditions do not lead to small
perturbations of the instanton geometry.  This can be understood in
a heuristic manner.  The CDL geometry has a compact throat
connecting its past and future regions. If we have an initial or
final state with too large an entropy, it will create a black hole
of radius larger than the throat.  This leads to a space-filling
space-like singularity, cutting the future off from the past.

The second proposal was to try to construct an analog of AdS/CFT
where the CFT lives on the boundary of the negatively curved
space-like slices of the CDL geometry. It is argued that the
appropriate boundary conditions for this situation allow quantum
fluctuations of the boundary geometry, so that the boundary CFT is
coupled to quantum gravity.  The hope is that this situation is well
defined when the boundary is two dimensional, and leads to a
boundary Liouville theory.   Two dimensional boundaries are
appropriate for 4 dimensional dS spaces, so this proposal relies on
the folk theorem that there are no dS solutions of SUGRA above 4
dimensions.

I do not understand the details of this construction or the
enthusiasm of its builders, so I will end this section with a list
of questions that I think must be answered, if this approach is
meaningful.

\begin{itemize}

\item What is the probability interpretation of the boundary field
theory? Only some of the extant theories of fluctuating two
geometries have a quantum mechanical interpretation. In those, the
genus expansion is the {\it divergent} $1/N$ expansion (actually the
double scaling limit) of a matrix quantum mechanics.  In this
context the genus expansion is said to converge.  What are the
probability amplitudes and what do they have to do with real world
measurements?  Is the theory quantum mechanics?  What are the
possible initial states?

\item Most of the asymptotically SUSic regions of moduli space are
decompactification limits, where the local asymptotically flat
space-time has dimension higher than four.   Why are only two
dimensional boundaries relevant?  One may want to argue that the
theory has a two dimensional boundary for all finite FRW times, but
the decompactifying dimensions should at least show up as an
infinite number of low dimension operators. The formalism has so far
restricted attention to massless bulk fields, but surely massive
fields whose mass asymptotes to zero must be part of the picture?

\item The construction is based on a particular instanton for decay
of a particular meta-stable dS point into a particular locally flat
geometry.  How do all the other instantons fit into the picture?
There must be some sort of monstrous duality in which the
observables are actually independent of the choice of instanton
geometry in the construction?

\item Conversely, how does one pick out of the Liouville/CFT observables, the
data relevant to our particular universe?  This is of course a
crucial step in trying to relate these ideas to the real world.  Is
there any relation between the answer to this question and the
practices of those landscape enthusiasts who simply do effective
field theory in a particular dS state?  Is the answer to this
problem computationally effective?  That is, can one really hope to
separate out the data corresponding to individual members of the
$10^{500}$ strong ensemble?

\item The construction purports to be a rigorous definition of what
is meant by the phrase {\it eternal inflation}.  What is its
prescription for the solution of the {\it measure problem} in that
context?  (Some progress has been made on the answer to this
question, but not enough to support phenomenological predictions).

\item The transition from a dS space with small positive c.c. and
one of the zero c.c. regions of the potential, is above the Great
Divide.  Supposedly one is saved from the problem of fluctuating
intelligent observers by much more rapid decays into negative c.c.
crunches. We are then left with the bizarre situation in which all
of the rigorously defined data about our universe can only be
measured in an extremely improbable history for the universe, one in
which it lasts long enough for all sorts of fluctuated intelligences
to exist.

\end{itemize}

I will not comment further on this proposal, except to mention that
I personally find the challenges of Holographic Space-Time and
Cosmological SUSY breaking much less daunting, and their connection
to actual observations infinitely more direct.  We turn next to an
explanation of the Holographic space-time formalism.

\section{Holographic space-time}
Having devoted much verbiage to the description of what a theory of
QG is {\it not} we are now ready to propose a general description of
what it is. This framework is meant to subsume all of the well
defined models we have discovered, which fall under the rubric
string/M-theory.  That claim has not yet been proven, and I will
admit from the beginning that a fully dynamical implementation of
the rules of {\it Holographic space-time} has not yet been found.

All well established models in the string/M-theory menagerie belong
to one of two classes. The first corresponds to space-times in
dimension $3 \leq d \leq 7$ with AdS asymptotics and an AdS
curvature radius that can be taken parametrically large, in the
sense that there is a closed set of boundary correlation functions,
which can be calculated in a systematic expansion about the
GKP/W\cite{gkpw} SUGRA limit. They all have exact AdS SUSY.

The observables in these models are correlation functions on a
boundary of the form $R \times S^{d-1}$. In addition, there are many
models of asymptotically flat space-time with dimension between
$4$\footnote{In 4 dimensions the complete theory of a gravitational
S-matrix, is complicated. Fadeev and Kulish\cite{fk} have given a
prescription, analogous to their treatment of electrodynamics. Not
much further work has been done exploring the consequences of their
ansatz.} and $11$. The only observable is the S-matrix. All of these
models have exact super-Poincare invariance.

In addition there are models which can be viewed as describing
certain infinite branes embedded in these spaces. In the AdS case,
these are relevant perturbations of the CFT describing the original
symmetric model\footnote{In calling these infinite branes, I am
working in the Poincare patch of AdS space, which corresponds to the
Hilbert space of CFT in Minkowski space. The corresponding solutions
in global coordinates are localized at the center of a global
coordinate system. There are also true brane solutions with AdS
asymptotics, analogous to D-branes embedded in flat space-time.} .
These models need not be supersymmetric, but they are
``supersymmetric in the majority of space-time". In the language of
CFT, this means that the high energy, short distance behavior is
dominated by a supersymmetric fixed point.  Although there are many
claims in the literature, there are no well established models with
nonsupersymmetric fixed points at large curvature radius.

We want to construct a more local description of QG, which will
reduce to these supersymmetric models in the infinite volume limit,
but which will enable us to describe systems that do not fall into
any of these categories, like cosmologies and the real world.  In
GR, local objects are never gauge invariant, so we should expect our
description to be adapted to a certain coordinate system. Indeed,
the fundamental postulates of the theory will contain in themselves
an explanation for why local physics can never be gauge invariant in
QG, a sort of quantum version of the principle of general
covariance.

The basic principles of holographic space-time are simple to state:

\begin{itemize}

\item The Strong Holographic Principle (Banks-Fischler) - A causal diamond is the
intersection of the interior of the backward light-cone of a point
$P$ with that of the forward light-cone of a point $Q$ in the causal
past of $P$. The boundary of a causal diamond is a null surface.
When we foliate it with space-like $d-2$ surfaces, we find one of
maximum area, called the holographic screen.  According to the
holographic principle, the quantum version of such a causal diamond
is a Hilbert space whose dimension is $e^{A\over {4 L_P^2}}$, where
$A$ is the area of the holographic screen. This formula is
asymptotic for large area. The proper quantum concept is the
dimension of the Hilbert space, which is of course always an
integer.

\item Intersections of causal diamonds correspond to common tensor
factors in the Hilbert spaces of two diamonds. Geometrically this
defines the area of the maximal causal diamond which fits in the
intersection.  Thus we have
$${\cal H}_1 = {\cal O}_{12} \otimes {\cal N}_1 $$
$${\cal H}_2 = {\cal O}_{12} \otimes {\cal N}_2 .$$ This encodes the
causal structure of the space-time, if we have a rich enough
collection of causal diamonds. We ensure this by beginning from a
lattice, which encodes the topology of an infinite space-like slice
(a Cauchy surface) of the manifold. For each lattice point ${\bf x}$
we have a sequence of Hilbert spaces ${\cal H} (n, {\bf x}) =
\otimes {\cal P}^n $, where ${\cal P}$ is a finite dimensional space
we will define below. Geometrically this represents a sequence of
causal diamonds whose future tips have larger and larger proper time
separation from the initial space-like slice.  For a model of a Big
Bang space-time we imagine the past tips to lie on the Big Bang
hypersurface. This incorporates the idea that the particle horizon
is very small near the singularity, but it is clear that nothing
singular happens in the quantum theory.  For a time-symmetric
space-time we take the lattice to lie on a time-symmetric space-like
slice, and the past and future tips of the diamonds lie an equal
proper time before and after the time-symmetric slice.

\item For nearest neighbor points on the lattice, at any $n$, we
insist that the overlap Hilbert space is $\otimes {\cal P}^{n-1} $.
We interpret these sequences of Hilbert spaces as the sequence of
causal diamonds of time-like observers, which penetrate the chosen
space-like slice at a given lattice point.  The proper time interval
between the tips of the $n$th diamond is a monotonically increasing
function of $n$. Thus, two nearest neighbor sequences of Hilbert
spaces, correspond to two time-like observers whose trajectories
through space-time are almost identical. The overlaps between other
points are constrained by two consistency conditions.   Let $d({\bf
x,y})$ denote the minimum number of lattice steps between two
points. We require that the overlap not increase as we follow a path
of increasing $d$, starting from ${\bf x}$, and that it decrease
asymptotically as $d({\bf x,y}) $ goes to infinity.

\item The second consistency condition is dynamical.  Let $N({\bf x})$ be the
maximal value of $n$ at a given lattice point.  We prescribe an
infinite sequence of unitary operators $U_k ({\bf x})$,operating in
the Hilbert space ${\cal H} (N({\bf x}), {\bf x})$, with the
property that for $k \leq N({\bf x})$ $U_k = I_k ({\bf x}) \otimes
O_k ({\bf x})$, where $I_k$ is a unitary in ${\cal H} (k, {\bf x})$
while $O_k$ operates in the tensor complement of this Hilbert space
in ${\cal H}(N({\bf x}), {\bf x}) $. This sequence is interpreted as
a sequence of {\it approximations to the S-matrix} in the time
symmetric case, and a sequence of {\it cosmological evolution
operators} in a Big Bang space-time.  We then encounter the
following set of fearsomely complicated consistency conditions.
Consider the overlap Hilbert space ${\cal O} (m, {\bf x}; n, {\bf
y})$. The individual time evolutions in ${\cal H}(N({\bf x}), {\bf
x})$ and ${\cal H}(N({\bf y}), {\bf y})$, each prescribe a sequence
of density matrices\footnote{There is no reason for the state on the
overlap to be pure. It is entangled with the other degrees of
freedom in each causal diamond.} on ${\cal O} (m, {\bf x}; n, {\bf
y})$. These two sequences must be conjugate to each other by a
sequence of unitary transformations.  A collection of Hilbert spaces
with prescribed overlaps, and evolution operators, satisfying all
the consistency conditions, is our definition of a quantum
space-time.

\end{itemize}

It's clear from this list, that any quantum space-time, which
approximates a Lorentzian manifold when all causal diamonds have
large area, will completely prescribe both the causal structure and
the conformal factor of the emergent geometry.  {\it We conclude
that in this formulation of QG, space-time geometry is not a
fluctuating quantum variable.} Given the results of
\cite{jacpadverl} it is likely that any geometry that emerges from
this framework will satisfy Einstein's equations with a stress
tensor obeying the dominant energy condition. This is because the
quantum system will obey the laws of thermodynamics, and those
authors claim that this is enough to guarantee Einstein's equations,
given the Bekenstein-Hawking connection between area and entropy.
Indeed, if we imagine {\it defining} the stress energy tensor as the
right hand side of Einstein's equations, then the only content of
those equations is whatever energy conditions we impose. The
holographic framework will certainly impose conditions sufficient to
prove the area theorem.

This observation is completely in accord with our semi-classical
conclusion that different asymptotic behaviors of space-time, even
if they are solutions to the same set of low energy field equations,
correspond to different models of QG.  The holographic construction
extends this principle to space-times whose boundaries are not
simple conformal transforms of static geometries.  One might object
that the standard Feynman diagram construction of perturbative QG
could not possibly be consistent with such a picture.  This is not
true.  These expansions only describe particles, including
gravitons, propagating in a fixed space-time background. Thus, to be
consistent with them, one must only require that the quantum
variables describe arbitrary scattering states of gravitons in
asymptotically flat or AdS space-times (or any other example over
which we claim to have good semi-classical control).

Our next task is to introduce just such variables.

\subsection{SUSY and the holographic screens}

Consider a ``pixel" on a holographic screen.  Naively, it's
characterized by a null vector and a plane transverse to it,
describing the orientation of this pixel in space-time. This is the
information content of solutions of the Cartan-Penrose equation
$$\bar{\psi} \gamma^{\mu} \psi (\gamma_{\mu})^{\alpha}_{\beta}
\psi^{\beta} = 0 ,$$ where $\psi$ is a commuting Dirac spinor.
Indeed, this equation implies that $n^{\mu} = \bar{\psi}
\gamma^{\mu} \psi $ is a null vector, and that $\psi$ itself is a
transverse or null-plane spinor corresponding to this null vector.
That is, if $\gamma^{\mu_1 \ldots \mu_k}$ are anti-symmetrized
products of Dirac matrices with $k \geq 2$ then
$$\bar{\psi}\gamma^{\mu_1 \ldots \mu_k}\psi $$ are non-zero only for
hyperplanes embedded in a particular $d-2$ plane transverse to
$n^{\mu}$.  The spinor has only $2^{[\frac{d-2}{2}]}$ independent
components. In eleven dimensions this is $16$ real components,
$S_a$.

The holographic principle implies that the Hilbert space of a pixel
should be finite dimensional, so the only operator algebra we can
write down for the $S_a$, consistent with transverse rotation
invariance, is

$$[S_a (n), S_b (n)]_+ = \delta_{ab} .$$  $n$ is a label for the
pixel, which we will discuss in a moment.  This algebra is the same
(up to normalization) as that of a massless superparticle with fixed
momentum in 11 dimensions. The smallest representation is the 11D
SUGRA multiplet, and all the others correspond to particles that,
according to the Coleman-Mandula theorem, cannot have an S-matrix
different from $1$.  If we think about different pixels, they should
have independent degrees of freedom, and we would normally ask that
the corresponding operators commute.  However each of the individual
pixel algebras has an automorphism $S_a (n) \rightarrow (- 1)^{F(n)}
S_a (n)$, which we treat as the $Z_2$ gauge symmetry called
$(-1)^F$. We can use this to choose a gauge where spinors
corresponding to different pixels anti-commute
$$[S_a (m) , S_b (n) ]_+ = \delta_{ab} \delta_{mn} .$$ The
spin-statistics connection familiar from local field theory is thus
built in to the holographic formalism.

Now let us think about the notion of pixel.  The holographic
principle again requires that a finite area holoscreen should have a
finite number of pixels, to each of which we assign a copy of the
single pixel algebra. The naive notion of pixel can be thought of as
a way to approximate the algebra of functions on the holographic
screen by the algebra of characteristic functions of a finite cover
of the screen by open sets. This opens the door to more general
approximations of the algebra of functions by finite dimensional
algebras that are not necessarily commutative. This has numerous
advantages.  For example, in the case relevant to the real world, a
two dimensional holographic screen with $SO(3)$ rotation invariance,
we can use the fact that $SU(2)$ has finite dimensional
representations of every integer dimension to construct the so
called fuzzy sphere.  The algebra of $N \times N$ matrices inherits
a natural action of $SU(2)$, which contains all integer spins
between zero and $N - 1$ .  It approximates the algebra of functions
on the sphere by the usual finite sums of spherical harmonics. The
specification of whether we get smooth, continuous, measurable or
square integrable functions is encoded in the behavior of the
expansion coefficients for large spin.

More generally, if the holographic screen has a Poisson structure,
there is a well developed theory of deformation quantization, which,
for compact manifolds, leads to a sequence of approximations to the
algebra of smooth functions by finite dimensional matrix algebras.
In general, this procedure has ambiguities; the analog of the usual
ordering ambiguities in quantum mechanics. However, for Kahler
manifolds there is much less ambiguity. The space of sections of a
holomorphic line bundle over a Kahler manifold is finite dimensional
and has a natural Hilbert space structure induced by the Kahler
potential. If we take sequences of holomorphic line bundles with
dimension going to infinity, we get natural fuzzy approximations to
the manifold.  Almost all of the manifolds that arise in string
compactification are related to Kahler manifolds in some way.
Calabi-Yau manifolds are an obvious example, and the Horava-Witten
bundles of Calabi-Yau manifolds over an interval are another. It is
not known whether general $G_2$ manifolds have a Poisson structure,
but those which exhibit non-abelian gauge groups, are $K3$
fibrations over a sphere or lens space.  A choice of Kahler form on
the $K3$, combined with the unique $SO(3)$ invariant Poisson
structure on $S^3$ or a lens space, defines a Poisson structure on
the entire 7-fold.

Combining these ideas, we obtain a general prescription for
compactification of holographic space-time. For compactifications to
4 dimensions we introduce variables satisfying the commutation
relations

$$[(\psi^M )_i^A , (\psi^{\dagger} N )^j_B ]_+ = \delta_i^j
\delta^A_B Z^{MN} \ \ \ i = 1 \ldots K, \ \ \ A = 1 \ldots K + 1.
$$

The operators $\psi$ and $\psi^{\dagger}$ are $K \times K + 1$ and
$K + 1 \times K$ matrices, sections of the two spinor bundles over
the fuzzy 2-sphere, the holographic screen for $4$ dimensional
space-time. The indices $M,N$ can be thought of as either minimal
spinor indices in $7$ dimensions or $(2,0)$ or $(1,1)$ spinors in
$6$. We know that in string compactifications with $8$ or more
supercharges, these different interpretations morph into each other
as we move around in moduli space. In the interior of moduli space,
where we expect the real world to lie, it may be that no particular
geometric description is picked out.  To be more precise, $M$ and
$N$ label a basis in the space of sections of the spinor bundle on
the appropriate manifold, appropriately truncated. This gives us a
possible new insight into string dualities.  It is well known for
example that the algebra of $N \times N$ matrices can actually be
thought of as a fuzzy approximation to the space of functions on
{\it any} Riemann surface. The topology and geometry of manifolds
emerges from fuzzy geometry in the large N limit, by discarding
different sets of matrices in the definition of the limiting
algebra.  In the interior of moduli space in string theory, where
the string coupling is not weak, compact manifolds have volumes that
are finite in Planck units and should therefore be thought of as
finite pixelations.  The dual geometry is obtained by taking a
different large N limit.

The operators $Z^{MN}$ are sums of $p-forms$ and we may think of
them as measuring the charges of branes wrapped around cycles of
manifolds.  More precisely, each $p-form$ component of $Z^{MN}$ will
be a sum of terms, each of which has such an interpretation.
Specifying the number of terms in this sum, for each $p$ will tell
us the number of independent $p-cycles$ in the manifold. In the
string theory literature, the $Z^{MN}$ are often called central
charges in the SUSY algebra. However we know that there are
interesting examples of singular manifolds, where their algebra is
non-abelian, and this gives rise to Yang-Mills gauge potentials in
the non-compact dimensions.

We have suppressed another set of matrix indices in the formula for
the anti-commutation relations above. Our internal spinors and
$p-forms$ are really sections of the corresponding bundles over some
fuzzy approximation to the internal manifold. The enumeration of
cycles in the previous paragraph is part of the structure of these
bundles. The geometry and topology of the manifold are all encoded
in the super-algebra of the generators $\psi, \psi^{\dagger} , Z$.
The smallest representation of this super-algebra, for fixed $i, A$,
is the pixel Hilbert space ${\cal P}$ referred to above.

An extremely interesting consequence of this method of
compactification is that fuzzy manifolds differ from each other
discretely.  There are no moduli.  This is a direct consequence of
the holographic principle and has nothing to do with dynamical
minimization of potentials.  We have noted above that space-time
geometry is part of the kinematical framework of holographic
space-time.  Our discussion of semi-classical gravity and the
principle that different solutions of the same gravitational field
equations can correspond to different quantum models, rather than
different states of the same model, here finds its ultimate
justification.   Continuous moduli can emerge from the holographic
framework when we take the dimension of the function algebra to
infinity.  There can be different ways to do this, and quantities
which go to infinity simultaneously at fixed ratio, define
continuous moduli of the limiting geometry.

For example, a fuzzy compactification of a Kahler manifold is
provided by the algebra of matrices in the space of holomorphic
sections of a line bundle over the manifold. The dimension of this
space is fixed by the element of the Picard group, which
characterizes the line bundle. These elements are labeled by
quantized $U(1)$ fluxes threading two cycles of the manifold and
(for ample bundles) the dimension goes to infinity along directions
in the Picard group where the fluxes go to infinity.  But there are
many such directions if the manifold has many two cycles, and the
ratios of fluxes through different cycles define continuous Kahler
moduli of the limiting manifold.

Note that one cannot really take this kind of limit for a single
pixel, or rather if one does so then one has taken the four
dimensional Planck length to zero. This would define, at best, a
free theory, analogous to free string theory, or at least an
interacting subsector that decouples from gravity. The moduli
problem of conventional string theory is a result of taking this
sort of limit as the starting point of the theory, and then
perturbing about it.  This remark is even more striking in the
context of the theory of stable de Sitter (dS) space that we present
in the next section.  It follows from the above remark, and the
assumption that this theory has a finite number of quantum states,
{\it that it has no moduli}. Furthermore, for a fixed value of the
c.c., the volume of the internal manifold in Planck units is
severely limited, and the limitation is related to the scale of SUSY
breaking!  We will deal with this in more detail below, but the
essential point is that the dimension of the Hilbert space of the
theory is $ \pi (RM_P)^2 = K (K + 1) {\rm ln D},$ where $D$ is the
dimension of ${\cal P}$, and $R$ the dS radius.  Using conventional
Kaluza-Klein ideas, we find ${\rm ln D} = (M_P / M_D)^2 = (V
M_D^{D-4})$. Here $D$ is either $10$ or $11$, $M_D$ is the $D$
dimensional Planck mass and $R$ the four dimensional dS radius. We
will see that the parameter that controls the validity of any four
dimensional effective field theory description is $K^{- 1/2}$. Thus,
a good field theory approximation, for fixed $RM_P$, requires $V
M_D^7$ to be bounded.

The key restriction on compactifications in this framework is that
the algebra of a single pixel should have a representation with
precisely one graviton and gravitino in the $K \rightarrow\infty$
limit. The classification of such algebras is one of the two central
goals of the holographic space-time program. The other is to find
equations that determine the scattering matrix.  By the way, our
focus on four dimensional compactifications is motivated by the
search for dS solutions of SUGRA.  In the limit $(\Lambda M_P^d)$
small, a quantum theory of dS space should produce a de Sitter
solution of a SUGRA theory.  The only known SUGRA Lagrangians that
have such solutions, and which also correspond to true
compactifications are Lagrangians with minimal SUSY in $d=4$. Such
Lagrangians can have many chiral multiplets, with a relatively
unconstrained Kahler potential and superpotential, which can easily
have dS minima.

\section{The theory of stable dS space}

The global geometry of dS space is described by the metric

$$ds^2 = - dt^2 + R^2 \cosh^2 (t/R) d\Omega_3^2 ,$$ where
$d\Omega_3^2$ is the metric on a unit $3$ sphere. As in
asymptotically flat or AdS spaces, we can obtain useful information
about the quantum theory by investigating perturbations, which do
not disturb the asymptotic behavior.  Since most ways of foliating
this geometry give compact spatial sections, the asymptotic regions
to be considered are past and future infinity.

To get an idea of the constraints on such perturbations, consider
the exercise of setting small masses $m$ on each point of the
sphere, {\it i.e.} making the ``co-moving observers" physical.  If
we do this at global time $T$, and space the masses by the
particle's Compton wavelength (since in a quantum theory, no
particle can be localized more precisely than that), then at $t= 0$
the particle number density is
$$m^3 \cosh^{3} (T/R), $$ and the $00$ component of the
stress tensor is exponentially large if $T \gg R$.  In other words,
long before $t = 0$, the back reaction on the geometry of the test
masses becomes important. In order to avoid this, we must make $m
\sim \cosh^{-1} (T/R)$ at time $T$.  This strongly suggests that,
{\it if we want to preserve dS asympotics in the future, we must not
try to fill the apparently huge volumes of space available in the
past with matter}.  Rigorous results along these lines have been
obtained in \cite{gary}\cite{boussofreivogel}.  The conclusion of
those studies is that if one inserts too much matter in the infinite
past, then a singularity forms before $t = 0$. If the singularity
can be confined within a marginally trapped surface of radius $<
3^{- \frac{1}{2}} R$, this can be viewed as a black hole excitation
of dS space, but if not, the whole space-time experiences a Big
Crunch and we are no longer within the class of asymptotically dS
space-times.

It is much simpler to understand the finite entropy of dS space, and
the arguments that this represents a Boltzmann counting of the total
number of quantum states corresponding to the thermodynamic
equilibrium state called ``the dS vacuum", from the point of view of
static coordinates, where

$$ds^2 = - d\tau^2 f(r) + \frac{dr^2}{f(r)} + r^2 d\Omega_2^2 ,$$
and
$$f(r) = (1 - \frac{R_S}{r } - \frac{r^2}{R^2}).$$ The parameter
$R_S d\equiv \frac{2M}{M_P^2}$ is the Schwarzschild radius of a
Schwarzschild-de Sitter black hole, and $R$ is the dS radius of
curvature. Empty dS space corresponds to $M = 0$. These coordinates
cover the maximal causal diamond of a time-like geodesic observer in
dS space.

Only the $\tau$ translation plus $SO(3)$ rotation generators
preserve the static coordinate patch.  If we consider quantum field
theory on the full dS manifold, then there is an action of the dS
group on the field theory Hilbert space, and for free fields, a
unique Gaussian state whose two point functions approach those of
the Minkowski vacuum at short distances. It has been known for a
long time that\cite{gibbonshawk} that this is a thermo-field state
for the thermal density matrix in the static patch\footnote{This is
a direct generalization of Israel's discussion\cite{Israel} of the
Hartle-Hawking vacuum in the Kruskal manifold.}, with temperature $T
= \frac{1}{2\pi R}$. Alternatively, this is the state which is
chosen by analytic continuation of Euclidean functional integrals on
the 4-sphere.

QFT in this geometry actually has an infinite number of states at
very low energy, where energy is defined as conjugate to the time
$\tau$, at $r=0$.  $f(r)$ vanishes near the horizon, $r = R$, so
there is a red shift of finite near horizon frequencies to low
frequencies at the origin.  If one uses the boundary conditions
imposed by the so called Bunch-Davies vacuum on the global dS
manifold, one finds an infinite number of states of arbitrarily low
energy.  It is important to realize that this is exactly the same
infinity encountered in global coordinates.  At $\tau = 0$ the
global geometry has only a finite size and all states are localized
in the causal diamond (the other half of the global geometry is just
a trick, the thermo-field double trick-for computing thermal
averages in the causal diamond).  As $\tau \rightarrow \infty$,
nothing falls through the horizon.  Rather things get pasted closer
and closer to the horizon and they redshift.

Within a causal diamond the infinity is analogous to the infinity of
near horizon states of a black hole. And, as in the black hole case,
there is a claim that the entropy of dS space is finite and equal to
one quarter of the horizon area in Planck units.  As with the black
hole, we must think of this entropy as representing the maximally
uncertain density matrix of the near horizon states, which means
that the number of states is finite.

Quantum field theory in a fixed space-time background encourages us
to think of dS space as having an infinite number of independent
horizon volumes, which are causally disconnected from each other.
The thermal entropy of a given horizon is interpreted as a finite
entanglement entropy between causally disconnected states of this
infinite system. This is supposed to explain the fact that the
entropy depends only on the area.  We have seen however that the
myth of independent horizon volumes is untenable because of
gravitational back reaction.  Our global considerations suggest a
total number of states for an eternal dS space, which is of order
the exponential of the Gibbons-Hawking entropy.

I will first outline some general properties of a theory of global
dS space, and then a more specific proposal, based on a cartoon of
the pixel algebra described in the previous section.  In my opinion
the correct theory will require us to understand the list of
consistent compactifications, which might be quite sparse.  It is
still within the realm of possibility that there is only {\it one}
consistent answer, and that it describes the real world.

\subsection{The two Hamiltonians of Wm. de Sitter}

Our theory of dS space has two Hamiltonians.  The first, $H$, has a
random spectrum, distributed in an interval of order $T =
\frac{1}{2\pi R}$.  Starting from a random initial state, that
Hamiltonian will generate expectation values for most operators,
which quickly become identical to their thermal averages in the
maximally uncertain density matrix.  The number of states in the
Hilbert space is of order $e^{\pi (RM_P)^2}$, and the average level
spacing is $T e^{-\pi (RM_P)^2}$. There will be recurrences on time
scales of order $R e^{\pi (RM_P)^2}$.

On time scales less than $ ~ R$, $H$ evolution will not make much of
a change in the state. We will postulate another Hamiltonian $P_0$,
which is useful for describing some of the states of the system over
these shorter time scales. $P_0$ will be the operator which
approaches the Hamiltonian of a super-Poincare invariant system in
the limit $RM_P \rightarrow\infty$. It will also be the appropriate
operator to identify with approximate descriptions of the system in
terms of quantum field theory in a background dS space\footnote{This
is somewhat confusing since that Hamiltonian is usually associated
with the static dS time coordinate. If we look at the action of the
corresponding vector field at some interior point of the static
observer's causal diamond, the static Hamiltonian converges to the
Poincare Hamiltonian.  However, they have very different actions on
the cosmological horizon. One should identify $H$ with the quantum
operator that implements static time translation on the horizon,
while $P_0$ is the corresponding action of the Poincare vector
field.}.

In order to understand it, we must first understand which states of
the system have such a field theoretic description. A local observer
can see only a region of physical size $R$, so we must ask how many
field theory like states can fit in such a region.  The density of
states of field theory in finite volume grows with energy and the
entropy of field theory states in a region of linear size $R$ is of
order
$$ (R M_c)^3 ,$$ where $M_c$ is the UV cutoff. The energy of a typical
state in this ensemble is
$$E \sim M_c^4 R^3 .$$ These estimates are valid as long as the
gravitational back reaction is small, a criterion which definitely
fails once the Schwarzschild radius $E/M_P^2$ is of order $R$. Thus,
we must have $$ M_c^4 R^2 < M_P^2 ,$$ which means that the entropy
is of order $ (RM_P)^{3/2} $, much less than the total dS entropy.
Most of the localized states in the horizon volume are black holes
whose radius scales like the horizon volume, and these states do not
have a field theoretic description in the horizon volume.

These estimates are valid for any low curvature region, and are
similar to the deficit between the entropy of a star and that of a
black hole of the same radius.  In dS space however we can interpret
the extra states as $(RM_P)^{1/2}$ copies of the field theoretic
degrees of freedom in a single horizon volume.  This allows us to
understand the picture of an infinite number of horizon volumes
predicted by QFT in curved space-time.  As with a black hole, one
should postulate a complementarity principle\cite{bhcomp}, according
to which the global description, is a description of the same system
as that in static coordinates. In the first the states are
interpreted as being localized in different regions, while in the
static coordinates the same set of states is seen as piled up at the
horizon.  The time evolution operators corresponding to the two
descriptions do not commute with each other.  In both the black hole
and dS systems, the holographic principle provides an infrared
cutoff on the number of states attributed to the system by QFT in
curved space-time.

Recall that the Schwarzschild-de Sitter metric is

$$ ds^2 = - dt^2 f(r) + \frac{dr^2}{f(r)} + r^2 d\Omega_2^2,$$ where
$ f(r) = 1 - \frac{R_S}{r} - \frac{r^2}{R^2} .$ The black hole mass
parameter is given by $ 2M = M_P^2 R_S$. This metric has two
horizons with $$R^2 = R_+^2 + R_-^2 + R_+ R_- $$ and $$ R_S R^2 =
R_+ R_- (R_+ + R_-) = R_+ R_- \sqrt{R^2 + R_+ R_-} .$$ Note that the
total entropy of this configuration {\it decreases} as the black
hole entropy $\pi (R_- M_P)^2$ {\it increases}. There is a maximal
black hole mass at which the Schwarzschild and cosmological horizon
radii coincide and equal $R_N = \frac{1}{\sqrt{3}} R$. The maximal
black hole is called the Nariai black hole.

This entropy formula suggests a model of the system in which the
Hilbert space has a finite number of states with logarithm $\pi (R
M_P)^2 $. Localized states are special low entropy configurations
with an entropy deficit, for small $R_S$\footnote{For general $M$,
the entropy deficit is $\Delta S = 2\pi R M (1 + \frac{\Delta
S}{S})^{-\frac{1}{2}},$ which indicates that large black holes are
present with somewhat larger than thermal probability.}
$$\Delta S = 2\pi R M .$$ If we interpret $M$ as the eigenvalue of a
Hamiltonian we will call $P_0$, this relation between the eigenvalue
and entropy deficit indicates that the maximally uncertain density
matrix is effectively a thermal distribution

$$\rho \propto  e^{ - 2\pi R P_0} ,$$ for eigenvalues of $P_0$
much less than the Nariai black hole mass. As a consequence, the
Poincare Hamiltonian, a generator acting on localized states in a
single cosmological horizon of dS space, which converges to the the
Hamiltonian of the super-Poincare invariant limiting theory when the
c.c. goes to zero, can be written
$$P_0 = \sum E_n  P_n .  $$  The $P_n$ are commuting orthogonal
projection operators, with $$ {\rm Tr}\ P_n = e^{\pi (RM_P )^2 -
\delta S_n }.$$ $\Delta S_n = 2\pi E_n R$, when $E_n \ll M_P^2 R$,
and near the maximal mass is given by the formula in the previous
footnote.

We can summarize the previous few paragraphs by saying that the
Bekenstein-Gibbons-Hawking formula for the entropy of black holes in
dS space motivates a model for the quantum theory of dS space in
which empty dS space is interpreted as the infinite temperature
ensemble of a random Hamiltonian $H$ bounded by something of order
the dS temperature ($||H|| \leq c T$)\footnote{The bound on the
Hamiltonian should be zero in the classical limit, consistent with
the classical notion of a vacuum.  This means it is of the form $T
f(T/M_P )$. Since the notion of localized observables in dS space
only makes sense when $\frac{T}{M_P} \ll 1$, the linear
approximation should be sufficient. So far I have not found any
measurable quantity whose value depends on $f(0) \equiv c $.  }.
This implies that localized black hole states are low entropy
deformations of the vacuum, and gives a connection between the black
hole mass parameter, which is the eigenvalue of another Hamiltonian
$P_0$, and the entropy deficit of its eigen-spaces.  This
observation leads us to expect what we already know to be true: {\it
the dS vacuum is a thermal state for quantum field theory with a
{\it unique temperature} $T = \frac{1}{2\pi R}$,} and the present
discussion can be viewed as an explanation of that fact from a more
fundamental point of view. It is particularly satisfying that this
explanation provides a rationale for the uniqueness of the dS
temperature.

There is another piece of semi-classical evidence that this picture
is valid. The Coleman-DeLucia formalism gives us an unambiguous
calculation of the transition rates between two different dS spaces.
As discussed above, the CDL formula implies that the ratio of the
two transition rates is given by the {\it infinite temperature limit
of the principle of detailed balance}. This is in perfect accord
with our model of the dS vacuum as the infinite temperature ensemble
in a Hilbert space of finite dimension. Similarly, Ginsparg and
Perry\cite{gp} and Bousso and Hawking\cite{bousshawk} have found
instantons for the nucleation of black holes in dS space, and their
results are completely consistent with the framework outlined above.

\subsection{Towards a mathematical theory of stable dS space}

It is my belief that the theory of dS space only makes sense in $4$
dimensions.  This follows from the basic principles I've enunciated,
plus a knowledge of low energy effective field theory.  The basic
principle we use is that SUSY is restored as the c.c. goes to zero,
with the gravitino mass going like $m_{3/2} = 10 K \Lambda^{1/4}$.
We will give two arguments for this behavior below.  This formula
implies that SUSY breaking must be describable in low energy field
theory, which in turn implies that it must be spontaneous, since the
gravitino mass and decay constant are much smaller than the Planck
scale. Supergravity Lagrangians in 5 or more dimensions do not have
de Sitter solutions\footnote{There are solutions of the form $dS
\times K$, where $K$ is a negatively curved manifold. If $K$ is
compact, there is no control over the amount of SUSY breaking,
because there are large corrections to the classical bulk solutions,
and both the dS and compact radii of curvature are naturally of
order the cutoff. For the dS radius this is just the fine tuning of
the c.c. in effective field theory, but the compact radius is an
additional fine tuning. Some of the literature considers non-compact
$K$, but throws away all but the constant mode on $K$. The meaning
of these papers is completely obscure to me. }, while in four
dimensions, models with chiral fields and appropriate super and
Kahler potentials can have lots of dS solutions.  This remains true
for dimension less than four. However, the interesting physics of dS
space is the localizable physics that is accessible to a local
time-like observer. As we will see, this is described by an
approximate S-matrix, which approaches that of a super-Poincare
invariant model as $RM_P \rightarrow\infty$. In $2$ and $3$
space-time dimensions there can be no such limiting theory, so there
is probably no useful model of low dimensional dS space either.

The notion of an approximate S-matrix can be formalized as follows.
Consider a causal diamond in dS space whose holographic screen has
an area $b \pi (RM_P)^2$, with $b < \frac{1}{2}$.  Assume also that
$(R M_P) \gg 1 $. According to the general principles of holographic
space-time there should be an {\it approximate scattering matrix}
$S(b, R)$, which operates on the eigenstates of the Poincare
Hamiltonian, relating two bases of eigenstates on the past and
future boundaries of the diamond. We do not yet have a prescription
for constructing $S$, but knowledge of effective field theory in $dS
$ space leads to the conclusion that this S-matrix becomes
insensitive to the $dS$ horizon as $R \rightarrow\infty$.

On an intuitive level this sounds obvious, but there is an important
subtlety.  We define the scattering matrix as the interaction
picture evolution operator $U(T, - T)$ in an effective field theory
in static coordinates.  The time $T$ is chosen such that the causal
diamond of the geodesic observer at the origin, between $- T$ and
$T$ has holoscreen area $b \pi (RM_P)^2$.  The intuitive argument
that this S-matrix becomes independent of $R$ as $RM_P \rightarrow
\infty$ is that the maximal Gibbons-Hawking temperature encountered
in that causal diamond is $ \frac{(1 - b)^{- \frac{1}{2}}}{2\pi R}$,
which goes to zero in the limit. The local geometry also approaches
Minkowski space. If we consider a configuration space Feynman
diagram contributing to the S-matrix, then all parts of it within
the causal diamond converge to their flat space values as the dS
radius goes to infinity.

As we approach the horizon, field theory in static coordinates
encounters an infinity.  The coefficient of $d\tau^2 $ vanishes,
which means that the norm of the Killing vector field
$\frac{\partial}{\partial\tau}$, goes to infinity.  As a
consequence, very high frequency modes of the field, localized near
the horizon, are low energy states as viewed from the origin. As we
approach the horizon, we appear to see an infinite number of modes,
all of which ``our friend at the origin" considers low energy.

A general relativist will attribute this to our insistence on using
"bad coordinates".  The message of the holographic principle is that
the pileup of states near the horizon is just the the holographic
image of all physical excitations which have fallen through the
horizon in coordinate systems that are regular there.  It also
instructs us to cut off the infinity, so that the total entropy of
these states is finite.  The latter instruction cannot be understood
in terms of quantum field theory, but must be built in to the
quantum theory of dS space we are trying to construct.  In the next
section, we will describe how thinking about Feynman diagrams with
internal lines that penetrate the horizon leads to a relation
between the gravitino mass and the c.c. .  We'll derive that
relation from different considerations in this section.

According to our general formalism, all the states in dS space are
accounted for in the irreducible representation of the pixel algebra
$$ [(\psi^M )_i^A , (\psi^{\dagger\ N})_B^j = \delta_i^j \delta^A_B
Z^{MN} , $$ where $M$ and $N$ run over a basis of sections of the
spinor bundle over the fuzzy compactification. For each pixel, the
irrep has dimension $D$ and we have
$$\pi (RM_P )^2 = K (K+1) {\rm ln\ } D . $$ In terms of Kaluza-Klein
language, ${\rm ln\ } D = V $, the volume of the compact dimensions
in higher dimensional Planck units. We also have the K-K relation $V
= (M_P / M_D)^2 $.

Particle states localized within our causal diamond are described by
considering the algebras of block diagonal matrices, with block
sizes $K_i$, with $\sum K_i = K$\cite{bfm}. The spinor bundle over
such an algebra is the direct sum of the set of $K_i \times K_i + 1$
matrices (and their $K_i + 1 \times K_i$ conjugates), each tensored
with the internal spinor bundle.  If, as $K \rightarrow\infty$, the
representation space of the pixel algebra approaches a direct sum of
supersymmetric particle state spaces, then the block diagonal
construction, with $K_i \rightarrow \infty$ and $\frac{K_i}{K_j}$
fixed, approaches the Fock space of that collection of
supermultiplets, with the correct Bose/Fermi gauge equivalence
(particle statistics).  We must of course include block
decompositions with an arbitrary number of blocks. Indeed, a direct
sum of algebras always has a permutation gauge symmetry, when we
view it as constructed from block diagonal matrices.

If $K$ is fixed and very large, only some of these block diagonal
constructions really resemble particles.  If $K_i$ is too small,
then the would-be particle will not be localizable on the
holographic screen, whereas if $K_i$ is too large there will not be
any multi-particle states.  The compromise, which maximizes the
entropy, while still retaining particle-like kinematics, is to take
each $K_i$ of order $\sqrt{K}$. The total entropy in such states is
of order $(RM_P)^{3/2}$, which is the same scaling we derived by
heuristic consideration of particle states in dS space, which do not
form black holes.

There are 3 important remarks to make about this construction.

\begin{itemize}

\item By considering off-diagonal bands in the block diagonalization
of the algebra of $K \times K$ matrices\footnote{This means the
$i$th upper off diagonal band, completed by the $K - i$th lower off
diagonal.}, we see of order $\sqrt{K}$ identical copies of the
highest entropy particle states.  These may be considered {\it
particle states in other horizon volumes} and we see how we can
reproduce the claim of QFT in curved space-time, in the
$K\rightarrow\infty$ limit\footnote{The transformations that map one
off diagonal band into the next should be thought of as discrete
analogs of the dS boosts, which change one static observer into
another.}. However, thinking in terms of the static coordinates, all
but one of these collections of particles should be lumped together
into the states on a particular observer's holographic screen. There
are of course of order $(RM_P)^2 $ such states.

\item The fixed ratios between the $K_i$ should be interpreted as
the ratios of magnitudes of the longitudinal momenta of the
different particles. Those familiar with Matrix Theory will
recognize this rule. We can motivate it by the following remarks.
The conformal group of the two sphere is the spin-Lorentz group $SL(2, C)$
and the spinor bundle contains solutions of the conformal
Killing spinor equation

$D_z s = \gamma_z s,$ where $z$ is a holomorphic coordinate on the
sphere and $\gamma_z$ is the pullback of the two dimensional Dirac
matrices by the zweibein. The solutions of the conformal Killing
spinor equation transform as a Dirac spinor $q^a_{\alpha}$ under
$SL(2,C)$. The requirement that the representations of the pixel
algebra are supermultiplets in the large $K_i$ limit implies in
particular that there are generators that converge to

$$S_a (\Omega_0 ) = S_a \delta (\Omega , \Omega_0) ,$$ where $S_a$
are two component real spinors under $SO(2)$, which satisfy a
Clifford algebra.  These operators are a ``basis" for the space of
sections of the spinor bundle.  They should be thought of as
operator valued measures on the space of sections. When we integrate
them against the conformal Killing spinors we get

$$Q_{\alpha} (\Omega_0) = \int\ S_a (\Omega_0 , \Omega) q^a_{\alpha}
(\Omega) = S_a q^a_{\alpha} (\Omega_0 ).$$ If $$[S_a , S_b ]_+ = p
\delta_{ab} ,$$ then
$$[Q_{\alpha}, Q_{\beta}] = (\gamma^0 \gamma^{\mu} )_{\alpha\beta}
P_{\mu} , $$ $$ P_{\mu} = p (1, \Omega) .$$ In deriving the
continuous generators from the fuzzy sphere, the normalization $p$
arises in the usual way.  The discrete generators differ from the
continuous one by an infinite normalization proportional to $K_i$,
so the ratios of $p_i$ are the ratios of $K_i$.

\end{itemize}

The precise super-particle spectrum that comes out in the limit
depends on the details of the rest of the pixel algebra
representation. The classification of pixel algebras whose limit
gives rise to a super-particle spectrum containing the $N=1$ SUGRA
multiplet is the analog in this formalism of classifying all
supersymmetric compactifications of string theory with minimal SUSY
in $4$ dimensions.  However, if we keep the pixel algebra fixed and
take $K \rightarrow \infty$, as is appropriate for a theory that is
the limit of stable dS space, then we only obtain models with no
moduli.   Other supersymmetric models, which can be described in
terms of perturbative string theory, come from more elaborate limits
in which we take both $K$ and the size of the pixel algebra to
infinity at the same time, obtaining continuous moduli. These are
not related to dS models.

The control parameter that governs the restoration of super-Poincare
symmetry is the typical particle momentum, $K_i$, which scales like
$\sqrt{K}$ . Rotational symmetry is of course exact, while the
Lorentz group is realized as the conformal group of the two sphere.
The accuracy with which it can be represented is limited by the
total number of spherical harmonics available, which scales like $K$
On the other hand, we can expect the violation of the super-Poincare
relation
               $$[Q_{\alpha} , P_0 ] \sim K^{- \frac{1}{2}} .$$
For a theory with spontaneously broken SUSY, the superpartner of any
state is that state plus one gravitino, so we get the estimate
$$m_{3/2} = K^{- \frac{1}{2}} M_P .$$  Taking into account the
relation between the dS entropy, $K$ and ${\rm ln}\ D$ we get

$$m_{3/2} = c {({\rm ln}\ D)}^{1/4} \Lambda^{1/4} = 10 c \Lambda^{1/4}.$$
The last estimate incorporates Witten's idea\cite{wittenuni} that
the volume of extra dimensions is the explanation for the ratio of
$100$ between the reduced Planck scale and the unification scale. We
might expect $c$ to be of order $1$ but we cannot say that we've
accounted for all factors of $2\pi$ correctly.   If $c$ is of order
$1$ then we get a gravitino mass of order $10^{-2}$ eV and a
gravitino decay constant $F \sim 30 ({\rm TeV})^2$.

\section{Implications for particle phenomenology}

I'll begin this section with an alternative derivation between the
gravitino mass and the cosmological constant, based on the notion of
Feynman diagrams with internal lines going through the horizon. We
want to consider a dS space with very large $RM_P$.  Low energy
physics is approximately the same as it is in the limiting
super-Poincare invariant model.  The latter is described by an $N=1$
SUGRA Lagrangian, with a super-Poincare invariant vacuum.  In order
to ensure that the cosmological constant is self-consistently zero,
we impose a discrete R symmetry on the low energy Lagrangian.  We
want to compute the leading correction to this supersymmetric
Lagrangian, which leads to the SUSY violation we expect in dS space.

This is computed, as effective Lagrangians always are, in terms of
Feynman diagrams, and the new effects of dS space obviously have to
do with diagrams in which internal lines go out to the horizon. They
cannot lead to explicit violation of SUSY, and renormalization of
parameters in the effective Lagrangian will not violate SUSY.
However, interactions with the horizon {\it can} violate R symmetry.
If we consider a diagram whose external legs are localized near the
origin, then lines going out to the horizon are extended over
space-like intervals of geodesic length $R$.  If we assume that the
gravitino is the lightest R charged particle in the model, the
leading R violating diagrams will have two gravitino lines leading
out to the horizon and will have an exponential suppression $e^{ - 2
m_{3/2} R}$.  It does not make sense to neglect the gravitino mass
in this formula, but the rest of the diagram is evaluated in the
$\Lambda = 0$ theory.   Recalling that the horizon has a huge number
(infinite in the field theory approximation) of very low energy
states, of order $e^{\pi (RM_P)^2}$, we can write the contribution
of this diagram as
$$ \delta {\cal L}\sim  e^{- 2 m_{3/2} R} \sum |< 3/2 | V | s >|^2 , $$
where $V$ is the operator representing emission from and absorption
of the gravitino by the horizon.

The horizon is a null surface and the massive gravitino can only
propagate near it for proper time of order its Compton wavelength.
As a quantum particle it does a random walk, and we take the proper
time step to be the Planck scale.  Thus, the area in Planck units
that it covers is of order $\frac{M_P}{m_{3/2}}$, and we take this
as an estimate of the logarithm of the number of states for which
the matrix element is of order $1$.   The total contribution is thus
of order

$$\delta {\cal L} e^{ - 2 m_{3/2} R + b\frac{M_P}{m_{3/2}}}.$$
This formula can be self consistent only for one behavior of the
vanishing gravitino mass in the $RM_P \rightarrow\infty$ limit.  If
we assume the gravitino mass goes to zero too rapidly, for example
like the naive SUGRA prediction $m_{3/2} \sim \sqrt{\Lambda}/m_P $,
then the formula predicts exponentially large corrections to the
effective Lagrangian.   If we assume it goes to zero too slowly the
effective Lagrangian is exponentially small, which is inconsistent
with the assumption.  In effective field theory, it is this
correction the the Lagrangian that is responsible for the gravitino
mass.   For self-consistency, the exponential dependence on $R$ must
cancel exactly
$$m_{3/2} \sqrt{\frac{b M_P}{2 R}} .$$  This is the same scaling we
found in the previous section, but we learn less about the
coefficient.

We conclude that the low energy Lagrangian of stable dS space has
the form
$${\cal L}_0 + {\cal L}_{\Delta R} .  $$ The full Lagrangian must
predict a dS solution, and implement the relation between the
gravitino mass and the c.c.  An example of such a Lagrangian would
be

$${\cal L}_{\Delta R} = \int\ d^2\theta\ (W_0 + F G),$$ with $G$ a
single chiral superfield, the goldstino multiplet, which we assume
is the only low energy matter field. ${\cal L}_0$ would have a
discrete $R$ symmetry, which forbade both of these terms. In order
that there be no SUSY vacuum in low energy effective field theory,
we have to assume that $G$ has R charge $0$. However, the demands of
the underlying theory are not so strict. We could for example insist
only that the R symmetry forbid terms up to cubic order in $G$ and
that the natural scale in ${\cal L}_0$ is just the Planck scale.
Then there might be SUSY minima at $S \sim m_P$, but the Lagrangian
could be above the Great Divide, and consistent with the underlying
finite dimensional model for dS space.

While this model satisfies the basic consistency conditions, it is
not our world.   In the real world, we must couple the SUSY
violating order parameter to standard model supermultiplets.  In
particular, gaugino masses would result from terms of the form

$$\int\ d^2 \theta\ f_i (G/M) W_{\alpha}^{i\ 2}  $$ and would be
given by
$$m_{1/2}^1 = f_i^{\prime} (G/M) (F/M) .  $$
Since $F \sim 30 ({\rm TeV})^2 , $ $M$ cannot be larger than a few
TeV if we are to obey the experimental bounds (there are factors of
standard model fine structure constants in $f_i$) .  This indicates
that there must be a new strongly coupled gauge theory with
confinement scale $M$, which contains fields transforming under the
standard model.  The Goldstino field $G$ must be an elementary field
with renormalizable couplings to the new gauge system, a composite
field from that system, or a combination of both.

Given these couplings, squarks and sleptons will get mass via gauge
mediation.  We are forced, by the low maximal scale of SUSY
violation, to consider a model of direct gauge mediation. Such
models are notorious for having problems with coupling unification.
One must have complete representations of the unified gauge group,
with low multiplicities, which means that the hidden sector gauge
group must be small and the representations of the new chiral matter
of low dimension.  So, for example, if the unified gauge group is
$SU(5)$, we can, when two loop corrections are taken into account,
tolerate at most $4$  $5 + \bar{5}$ pairs in the hidden sector.

While I have not done a definitive survey, all examples I've studied
of hidden sectors that satisfy these constraints contain light
fields with standard model quantum numbers, which are ruled out by
experiment.  Simple unification appears incompatible with direct
gauge mediation. One appears forced to utilize Glashow's {\it
trinification} scheme, in which the standard model is embedded in
$$SU_1 (3) \times SU_2 (3) \times SU_3 (3) \rtimes Z_3 ,$$ where
$Z_3$ cyclically permutes the three $SU(3)$ groups.  The standard
model chiral superfields are embedded in $3$ copies of $$
(1,\bar{3}, 3) + (3,1,\bar{3}) + (\bar{3},3,1), $$ as the $15$
states that transform chirally under the standard model.  There is a
nice embedding of this in $E_6$, but that would put us back in the
forbidden realm of simple unification. More interesting is the way
that this structure, {\it including the prediction of the number of
generations}, arises from $3$ D3-branes at the $Z_3$ orbifold in
Type IIB string theory. We also note that the vector-like spectrum
of this model contains $3$ copies of the conventional SUSY Higgs
fields.  However, at least in the orbifold construction the implied
structure of standard model Yukawa couplings comes out wrong.

We can add a hidden sector to trinification, without ruining
standard model coupling unification, by postulating an $SU(N)$ gauge
theory, with $N = 3,4$ and chiral fields $T_i$ and $\tilde{T}_i$ in
the $(\bar{N}, 3_i) + (\bar{3}_i , N) $. These models have a
pyramidal quiver diagram and are called the Pyramid
Schemes\cite{pyrma}.  At the level of the orbifold construction the
new fields come from $D7$-branes and one can think of the model as
an F-theory solution with an orbifold singularity in its base.

There is no room here to go into the intricate details of model
building, but the Pyramid Schemes throw new light on the strong CP
problem, the little hierarchy problem, the origin of the $\mu$ term
in the MSSM, the nature of dark matter, {\it etc.}.  They have a
rich phenomenology and can easily be ruled out at the LHC.  It is
not clear whether the LHC energy is high enough to reveal the
complete structure of these models.

What I would like to emphasize is that the theory of stable dS space
we have adumbrated gives rather detailed predictions for Terascale
physics.  Thus, despite its rather abstract origins, and the
incomplete nature of the theory of holographic space-time, we may
hope in the near future for experimental input that could encourage
us to continue to work on this set of ideas, or convince us to
abandon them.

\vfill\eject

\section{Appendix: exercises on CDL Tunneling}

 In my lectures, I asked the students to work out some of the theory
 of gravitational tunneling for themselves, because there are so
 many erroneous notions in the community about the results of
 Coleman and De Luccia.  Much of my second lecture was an extended
 ``recitation section", in which I outlined the solution of these
 problems.  There are also some exercises on black hole solutions.

\noindent 1. Show that the metric

$$ ds^2 = - f(r) dt^2 + {{dr^2}\over f(r)} + r^2 d\Omega^2 , $$
with $$f(r) = 1 - c_d {{M }\over {r^{d-3} M_P^{d-2}}} \pm
{{r^2}\over {R^2}} ,$$ solves the $d$ space-time dimensional
Einstein equations with cosmological constant $\Lambda$ , where
$R^{-1} = b_d  \sqrt{\Lambda} / M_P^{d/2 - 1 } $ is the Hubble scale
associated with the c.c. .   Work out the necessary constants for
all $d$.   The Einstein equations are

Show that positive c.c. corresponds to the choice of negative sign
in $f(r)$. These are the Schwarzschild black hole solutions for all
possible maximally symmetric background space times.

\vskip.2in

\noindent 2. Show that for positive c.c. $f(r)$ has two zeroes,
corresponding to the two positive roots of a cubic equation
$$(r - R_+) (r - R_-)(r + R_+ + R_-).$$  $R_{\mp}$
is the position of the black hole (cosmological) horizon.  Show that
both $R_{pm} $ are $< R$ and the entropy deficit $$\pi  (R^2 - R_+^2
- R_-^2) M_P^2 $$ is always positive and is approximately $$\Delta S
\simeq 2\pi R M$$ when $R_- << R$. Find the maximal black hole mass
in de Sitter space and argue that it has the smallest total entropy.
This little exercise shows that localized states are low entropy
excitations of the dS vacuum, which we have argued should be modeled
by an infinite temperature density matrix on a Hilbert space with a
finite number of states.

\noindent 3. The Coleman De Lucia (CDL) equations for gravitational
tunneling are the equations for a scalar field coupled to Euclidean
Einstein gravity, with $SO(4)$ symmetry.  This is the Euclidean
analog of FRW cosmology: a four dimensional space-time with a
maximally symmetric 3 dimensional subspace.  The equations are

$$ \phi^{\prime\prime} + 3 {{\rho^{\prime}}\over \rho} \phi^{\prime} =
{{dU}\over {d\phi}} ,$$ $$  {\rho^{\prime}}2 = 1 +  {{\rho^2}\over
{3 m_P^2}}({1\over 2} \phi^{\prime\ 2} - U) .$$   The metric is
$$ds^2 = dz^2 + \rho^2 (z) d\Omega_3^2 .$$

\noindent A.If the potential has the form $$U = - \mu^4 v(x),$$
where $x = \phi / M$, and we make everything dimensionless using the
space-time scale $z = {M\over {\mu^2}}\tau$, show that the equations
take the form
$$\ddot{x} + {{\dot{a}}\over a} \dot{x} = - {{dv}\over {dx}},$$
$$\dot{a} = 1 + \epsilon^2 a^2 ({1\over 2} \dot{x}^2 + v),$$
where $\epsilon^2 = {{M^2}\over {3 m_P^2}} .$

If we are tunneling {\it from} a solution with non-positive
c.c.,then the Euclidean space-time is infinite.   We insist that, at
infinity, the solution approach the Flat or Hyperbolic space
solution.  By convention we set the field value at infinity to zero,
and $v(0) = c$, so that the c.c. is $- c \mu^4 $.  The other maximum
of $v$ is called $x_T$. The solution for the metric at infinity is
$$a = \sinh (\epsilon c \tau ) / \epsilon c \rightarrow \tau .$$
The last limit is $c \rightarrow 0$. The exact solution will also
have a point where $a = 0$.  This is the center of the vacuum
bubble.

The equations (but not the metric) have the form of an FRW {\it
EUniverse} with a scalar field. The ``Big Bang" is the center of the
bubble, which we conventionally call $\tau = 0$ and we must have
$\dot{x} = 0 $ there in order to have a regular solution.  We must
choose the value of $x(0)$ in order to satisfy the boundary
conditions at infinity.   We choose $\dot{a} (0) = 1$.  This
expanding Euniverse condition just says that we are following the
Euclidean configuration to larger radius spheres. In Euclidean
space, this analog Big Bang is not a singularity. Since $c \geq 0$,
the real, Lorentzian signature c.c. is non-positive, but the analog
EUniverse has non-negative c.c. at the maximum of $v$ at $x = 0$.
The equations correspond to motion under a complicated frictional
force, plus the force derived from the potential $v$. Show that the
EEnergy, $$EE = \frac{1}{2} \dot{x}^2 + v , $$ will be monotonically
decreasing as long as $\dot{a}$ remains positive, as will the speed.

\noindent B. In the non-gravitational case ($\epsilon = 0$)
$\dot{a}$ is always positive.  Then, it's clear that the friction
term goes to zero at large $t$.  Argue that there are values of
$x(0)$, such that $x(t)$ will undershoot the maximum of $v$ at $x =
0$.  Argue that by starting close to $x_T$ we can find solutions
that overshoot $x=0$. Argue that continuity implies there is a
solution, which asymptotes to $0$ at $\tau = \infty$.   Argue that
this analysis remains valid when $\epsilon \ll 1$, as long as $v$ is
of order one and $v(x_T) - v(0)$ is of order one.

\noindent C.  When $\epsilon$ is of order $1$ a new behavior sets
in.  The sign of $\dot{a}$ can change in the region where $v$ is
negative. If this happens, the solution never reaches infinite
radius. The radius shrinks to zero and the $\dot{x}$ does not go to
zero.  Friction turns to anti-friction as the radius shrinks and the
velocity actually goes to infinity.  We can think of this as a Big
Crunch of the EUniverse. Such solutions exist when $\epsilon$ is
very small, but they are confined to a small range of values of
$x(0)$ near the minimum of $v$. The transition between overshooting
and undershooting happens at a larger negative value of $x (0)$ so
the instanton always exists. Argue that as $\epsilon$ is increased,
the onset of Big Crunch solutions moves to more negative values of
$x$.  Eventually it crosses the transition point between overshoot
and undershoot solutions, and no instanton exists.

\noindent D. When $c > 0$, even when the instanton exists, its
interpretation is not that of an unstable bubble that can appear as
a state in the AdS space.  Argue this as follows:  Near infinity the
instanton $x(\tau )$ becomes small, and is well approximated by a
solution of the equations for small fluctuations around Euclidean
AdS space.   Argue that, because of the boundary conditions on the
instanton at $\tau = 0$ it is a linear combination of both the
normalizable and non-normalizable solutions of the linearized
equations. As you will learn in other lectures on the AdS/CFT
correspondence, this means that it corresponds to adding an operator
to the Hamiltonian.  States in the model with the original
Hamiltonian correspond to purely normalizable solutions at infinity.
In all cases of the AdS/CFT correspondence where such instantons
have been found, the operator that is added is unbounded from below.

\noindent E. The overshoot solutions are those for which $x(0)$ is
near $x_T$. Thus, as $\epsilon$ is raised, the point at which
instantons disappear is the point at which $x(0)$ is forced to $x_T$
in order to avoid a crunch.  However, this is no longer an
instanton, because if we start a solution at $x_T$ with zero
velocity, it stays there. What happens instead is that the point
recedes in geodesic distance, and the solution becomes infinite in
both the $\tau = 0$ and $\tau = \infty $ limits.  Show that the
interpretation of this solution is as a static domain wall between
two AdS regions (or an AdS and Minkowski region).   Show more
generally that the existence of such a static domain wall always
requires the fine tuning of one parameter in the potential.  We
summarize this in the statement that the sub-manifold in the space
of potentials, on which a static domain wall solution exists, {\it
has co-dimension one}.  This sub-manifold is called The Great
Divide. On one side of the Great Divide instantons exist, while on
the side we call Above the Great Divide, they don't exist.  There is
a connection between this and the positive energy theorem in General
Relativity, which I will explain in the lectures.

\noindent F. Show that the Euclidean continuation of dS space is a 4
sphere, and that it has negative Euclidean action. In fact, in an
echo of the Gibbons-Hawking\cite{GH} result for Euclidean black
holes, the action is just equal to minus the entropy of dS space.
Correspondingly, instantons for the ``decay" of dS are compact $4$
manifolds with negative Euclidean action.  We make a probability
formula that is $\leq 1$ by subtracting the negative dS
action\footnote{The rule of subtracting off the action of the
initial configuration is motivated by quantum field theory, where we
can {\it prove} that this is the right thing to do.}  of the initial
"decaying" state.
$$ P_{12} = e^{- (S_I - S_{dS_1}} .$$  If the state $2$ to which
$dS_1$ "decays" is also a dS space, then we can form the reverse
probability

$$P_{21} = e^{- (S_I - S_{dS_2})} .$$   This leads to

$$\frac{P_{12}}{P_{21}} = e^{- \Delta\ Entropy}.  $$ Argue that this
is the infinite temperature form of the {\it principle of detailed
balance}.  Consider two finite collections of states, such that the
transition amplitude for any state in collection 1 to any state in
collection 2 is the same.   Show that unitarity implies that there
are reverse transitions, and that the probabilities for the two
ensembles to decay into each other are related by the above
equation.   The CDL formula thus provides evidence for the picture
expounded in the lectures, in which dS space is modeled as a system
with a finite number of states.  Notice that the instanton
transition for the lower c.c. dS space, is here interpreted, not as
an instability, but as a temporary sojourn of a large system in a
very low entropy configuration, like the air in a room collecting in
a little cube in the corner.

More controversial is the contention, also expounded in the
lectures, that the same interpretation is valid {\it above The Great
Divide} for dS ``decays" into negative c.c. Big Crunches.  The
holographic principle shows that the latter are low entropy states,
and we should expect rapid transitions back from them to the
equilibrium dS configuration.  These reverse transitions, cannot be
modeled by instantons, because the initial configuration is not
classical in any way.

\section{Appendix: potentials in string theory}

In tree level string theory, one can only add sources to the system
if they correspond to vertex operators for asymptotic states of the
system in a fixed space-time background. In asymptotically flat
space, this means that one can only add constant sources, as in the
definition of the field theoretic 1PI potential, for massless
particles. They correspond to rather singular limits of genuine
scattering amplitudes, but, {\it so long as the particle remains
massless for all values of the source}, they seem sensible.  The
italicized phrase means that the analog of the effective potential
can only be defined when it is exactly zero.

In non-supersymmetric string theory, even when there are no
tachyons, the perturbation expansion is singular at one loop.
Fischler and Susskind\cite{fs} argued that these singularities could
be removed by changing the background space-time.  This procedure
leads to time dependent solutions, and general considerations
show\cite{bdcosmo} that the time dependence is singular\footnote{In
a systematic F-S expansion, the time dependence appears linear but
at large times this expansion breaks down.  One can try to do a more
exact solution of the low energy field equations, but this leads to
singular cosmological solutions. There does not appear to be a way
to make the F-S mechanism into a controlled expansion.}.

Fischler and Susskind tried to argue that their procedure gave a
method for computing quantum corrections to the effective potential
in string theory.  They showed that there was a Lagrangian, at the
appropriate order in string coupling, which reproduced the modified
background solution they had found.  Students who have studied the
rest of these lectures, will know that such a demonstration says
nothing whatever about the existence of other solutions of the same
equations of motion, as bona fide theories of quantum gravity.  This
argument is independent of the question dealt with above, to the
effect that the Fischler Susskind solution itself does not provide
evidence for the existence of a model of quantum gravity based on
their modified background.

Another attempt to define effective potentials in string theory
tries to define a String Field Theory\cite{strfldthry}.  Open String
Field Theory is an elegant construction, which reproduces tree-level
open string amplitudes.  However, at the loop level it is singular,
because of the familiar fact that open string loops imply closed
strings.  {\it Any} regularization of that singularity forces us to
introduce an independent closed string field.  Closed String Field
Theory is {\it not} a non-perturbative definition of theory.  Its
Lagrangian must be corrected at each order in perturbation theory,
in order to reproduce the correct loop amplitudes. Furthermore, the
series that defines the string field action is divergent.  Much has
been made of the fact that the open string field theory ``contains
closed strings automatically", and it's been proposed that this
gives a non-perturbative definition of the theory.  In fact, the
appearance of closed strings is ambiguous and the relevant open
string diagrams are singular.   When one tries to regulate the
singularities, one finds that one must introduce an independent
closed string field, with the difficulties noted above.

The upshot of this is that there is no indication in any
perturbative string theory calculation, that there is a beast like
the mythical effective potential, whose minima classify different
consistent theories of quantum gravity.  Every non-perturbative
definition of string theory leads to precisely the opposite
conclusion, as we have sketched in the main lectures.



\end{document}